\documentclass[aps,prb,twocolumn,superscriptaddress,floatfix]{revtex4}

\usepackage[vcentermath]{youngtab}
\usepackage{amscd,rotating}
\usepackage{amsmath} 
\usepackage{amssymb} 
\usepackage{graphicx}     
\usepackage{longtable}
 
\begin{document}
 
\newcommand{\nin}{\ensuremath{\not\in}}
\newcommand{\bfc}{\mathbf{c}}
\newcommand{\rmd}{\mbox{d}}

\title{Transition from a strong-coupling fixed point to an 
intermediate-coupling fixed point in a single-channel SU(N) Kondo model: 
role of the filling and two-stage screening}


\author{Andr\'es Jerez}

\affiliation{European Synchrotron Radiation Facility, 6, rue Jules
Horowitz, 38043 Grenoble Cedex 9, France}

\author{Mireille Lavagna}

\altaffiliation{Also at the Centre National de la Recherche Scientifique
(CNRS).}

\affiliation{Commissariat \`{a} l'Energie Atomique, DRFMC /SPSMS, 17,
rue des Martyrs, 38054 Grenoble Cedex 9, France
}

\author{Damien Bensimon}

\affiliation{Commissariat \`{a} l'Energie Atomique, DRFMC /SPSMS, 17,
rue des Martyrs, 38054 Grenoble Cedex 9, France
}

\affiliation{Department of Applied Physics,
University of Tokyo, Bunkyo-Ku,
TOKYO 113-8656
JAPAN
}

\date{\today}

\begin{abstract}
We study an extended SU(N) single-impurity Kondo model in
which the impurity spin is described by a combination of
Abrikosov fermions and Schwinger bosons. Our aim is to describe
both the quasiparticle-like excitations and the locally critical
modes observed in various physical situations, including
non-Fermi liquid behavior in heavy fermion systems in the
vicinity of a quantum critical point. 
We identify the strong coupling fixed
point of the model and study its stability within second order perturbation
theory. Already in the single channel case and in contrast with
either the pure bosonic or the pure fermionic case, the
strong coupling fixed point is unstable against the conduction
electron kinetic term as soon as the amount of Abrikosov fermions
reaches a critical value. 
In the stability region, the partially screened, dressed impurity
at site 0 repels the conduction electrons on adjacent sites. In the instability
region, the impurity tends to attract $(N-1)$ conduction electrons to the 
neighboring sites, giving rise to a two-stage Kondo effect with 
additional screening.
This result opens the route to the existence
of an intermediate coupling fixed point, characterized by non-Fermi liquid 
behavior.  
\end{abstract} 

\maketitle


\section{Introduction}

Recent experiments in Heavy-Fermion compounds have shown the existence of a
quantum phase transition from a magnetically disordered to a long-range
magnetic ordered phase, driven by change in chemical composition, pressure
or magnetic field \cite{sachdev}. 
For an extensive survey of the 
experimental situation we refer the reader to the review article of 
Stewart \cite{stewart}. 
In a very unusual way, the behavior of the system in the
disordered phase close to the quantum critical point (QCP) differs from that
of a Fermi-liquid. For example $CeCu_{6-x}Au_{x}$ \cite{vl94,vl96} and 
($Ce_{1-x}La_{x})Ru_{2}Si_{2}$ \cite{kambe96} present an antiferromagnetic transition,
respectively, at $x_{_{C}}=0.1$ and at $x_{_{C}}=0.08$. While far from the
QCP, the magnetically disordered phase is a Fermi liquid
with a large effective mass, the temperature dependence of the physical
quantities in the disordered phase in the vicinity of the QCP is of
non-Fermi-liquid like type. Typically, in $CeCu_{5.9}Au_{0.1}$ \cite{vl96}, the specific
heat $C$ depends on $T$ as $C/T\sim -ln(T/T_{0})$, the magnetic
susceptibility as $\chi \sim 1-\alpha \sqrt{T}$, and the $T$-dependent part
of the resistivity as $\Delta \rho \sim T$ instead of $C/T\sim \chi \sim
Const$ and $\Delta \rho \sim T^{2}$ as in the Fermi-liquid state. Once a
long-range magnetic order is set up, the effect of a pressure or of a
magnetic field is to drive the system back to a magnetically disordered
phase with a Fermi-liquid behavior. The same type of behavior has been
observed in other systems such as $YbRh_{2}Si_{2}$ \cite{trovarelli00}, 
$CeNi_{2}Ge_{2}$ \cite{steglich97}, 
$CeCu_{2}(Si_{1-x}Ge_{x})_{2}$ \cite{steglich96}, $CeIn_{3}$, $CePd_{2}Si_{2}$
\cite{mathur98}, and $U_{1-x}Y_{x}Pd_{3}$ \cite{seaman91}. 
The associated breakdown of the
Fermi-liquid theory poses fundamental questions about the possible formation
of novel electronic states of matter with new types of elementary
excitations resulting from the presence of strong correlations among
electrons.

On the theoretical side, two scenarios are in competition to describe
quantum phase transitions: either the itinerant magnetism scenario (i), or\
more recently proposed, the locally critical picture (ii).\ 
\smallskip 

In the former case, (i), the quasiparticles still exist at the QCP and the
theory focuses on the study of the low-lying, large-wavelength (low-$\omega $,
low-$q$) fluctuations of the order parameter close to the transition. The
calculations have been performed within the renormalization group scheme
\cite{hertz76,millis93,continentino93} or in the self-consistent spin-fluctuation
theory \cite{moriya95}, and have been recently extended \cite{lp00} to the microscopic model
which is believed to describe the Heavy Fermions, the Kondo lattice. In all
the cases, they lead to a $\Phi ^{4}$-theory with an effective dimension $%
d_{ef\!f}=d+z$ where $d$ is the spatial dimension and $z$ is the dynamic
exponent.\ In the experimental situations, $d_{ef\!f}$ is above its upper
critical value equal to 4, since $d$ is equal to 3 or 2, and z varies 
from 2 to 3
depending whether the spin fluctuations are staggered or uniform. Hence the
system is described by a Gaussian fixed point with anomalous temperature
dependence of $C/T$ and $a=\Delta \rho /T$ but with predictions which cannot
account for the non-Fermi-liquid behavior observed experimentally. 
\smallskip 

The second scenario, (ii), has been motivated by the
recent results obtained by inelastic neutron scattering experiments
performed on $CeCu_{5.9}Au_{0.1}$. The dynamical spin susceptibility 
$\chi^{\prime\prime}(\mathbf{q},\omega )$ near the magnetic instability 
wavevector $\mathbf{Q}$
has been found to obey an anomalous $\omega /T$ scaling law \cite{schroder98,schroder00} 
as a function of temperature: 
$\chi^{\prime\prime}(\mathbf{Q},\omega )\sim 
T^{-\alpha }g(\omega /T)$ with
an exponent $\alpha $ of order $0.75$.\ Moreover, such a $\omega $ and $T$
dependence appear to stand over the entire Brillouin zone revealing in the
bulk susceptibility too. This fact strongly suggests that the spin dynamics
are critical not only at large length scales \cite{coleman99,si01} but also at atomic length
scales contrary to what happens in the traditional itinerant magnetism
picture, (i). From these results, one can deduce that local critical modes
coexist with large-wavelength fluctuations of the order parameter implying
 a non-Gaussian fixed point beyond the $\Phi ^{4}$-theory. Alternative theories
to the spin-fluctuation scheme are needed to describe the local feature of
the quantum critical point characterized by the simultaneous disappearance
of the quasiparticles and the formation of local moments. In this direction, we
will mention recent calculations based on a dynamical mean field theory \cite{gremp,burd}
(DMFT) which seem to lead to encouraging results concerning the scaling law
variation of $\chi^{\prime\prime}(\mathbf{q},\omega )$. 
\smallskip

A recent approach which has been
developed in order to describe the local QCP, is based on a supersymmetric
theory \cite{gca92,pl99,cpt1,cpt2} in which the spin is described in a mixed fermionic-bosonic
representation. The interest of the\ supersymmetric approach is to describe
the quasiparticles and the local moments on an equal footing through the
fermionic and the bosonic part of the spin, respectively. It appears to be
specially well-indicated in the case of the locally critical scenario in
which the magnetic temperature scale $T_{N}$, and the Fermi scale $T_{K}$
(the Kondo temperature) below which the quasiparticles die, vanish at the
same point, $\delta _{C}$.
\smallskip

An important aspect in the discussion of the breakdown of the
Fermi-liquid theory is related to the question of the stability of the
strong coupling (SC) fixed point. 
Whereas all the issues presented previously concerning Heavy fermion systems  
have to do with properties of the lattice, the instability of the SC fixed 
point can be regarded already by studying the single impurity problem.

The traditional source of instability 
in the single impurity Kondo model is the presence of several
channels for the conduction electrons with the existence of two regimes, 
underscreened and the overscreened, with very different behaviors as we 
are about to recall. Indeed we will see that
this is not the only possible source of instability of the strong coupling
fixed point. Recent works have shown that more general Kondo impurities of
symmetry group $SU(N)$ may also lead to an instability of the SC fixed point
already with one channel of conduction electrons. 

In order to fix ideas, let us start with the antiferromagnetic 
single-channel Kondo impurity model.
It is well known that within a
renormalization group (RG) analysis \cite{pwa67,pn76,hew}, the flow takes the Kondo coupling J all the
way to strong coupling. The weak-coupling beta function follows the
renormalization group equation
\begin{equation}
\beta (g)=\frac{dg(\Lambda )}{d\Lambda }=-g^{2} \label{betaf}
\end{equation}%
where $g=\rho _{0}J$ and $\rho _{0}$ is the density of states of conduction
electrons. The system flows a to strong coupling fixed point which is stable
and the associated behavior of the system is that of a local Fermi liquid. 

The situation is rather different when one considers several 
channels for the conduction electrons.\ In
the case of a spin $S$ of symmetry group $SU(2)$ in Kondo interaction with
conduction electrons belonging to $K$ different channels, Blandin and Nozi\`{e}res
\cite{nb} have shown that the multichannel Kondo model can lead to two very
different situations depending how $K$ compares to $2S$.\ Their calculation
corresponds to a second order perturbation theory in the hopping
amplitude, $t$, of the conduction electrons, around the strong coupling
fixed point.
They analyze their results by deriving an effective coupling, $J_{ef\!f}$ between
the spin of the composite formed by the impurity dressed by the conduction
electrons in the strong coupling limit, and the spin of the conduction
electron on the neighboring sites. They are then able to apply the same RG
analysis to $J_{ef\!f}$ as indicated in Eq. (\ref{betaf}). 
In the underscreened regime,
when $K<2S$, the effective coupling is found to be ferromagnetic and the
strong coupling fixed point is stable. In the overscreened regime when $K>2S$%
, the effective coupling is found to be antiferromagnetic and hence the
strong coupling fixed point is unstable. The former $K<2S$ regime
corresponds to the one-stage Kondo effect with the formation of an effective
spin, $(S-1/2)$, resulting from the screening of the impurity spin by the
conduction electrons located on the same site. The system described by the
strong coupling fixed point, behaves as a local Fermi liquid. The
instability of the strong coupling fixed point obtained in the latter regime, 
$K>2S$, is associated with a multi-stage Kondo effect in which successively
the impurity spin is screened by conduction electrons on the same site, and
then the resulting dressed impurity\ is screened again by conduction electrons 
on the
neighboring site and so forth.
The instability of the strong coupling fixed point in the
oversreened regime is an indication of the existence of an intermediate
coupling fixed point which has been then investigated \cite{ad84,wt83,cr93} by means of other
methods. As it is well established now, the intermediate coupling fixed
point leads to non-Fermi-liquid excitation spectrum with an 
anomalous residual entropy at zero temperature. 
\smallskip

It has recently put forward that other sources of instability of the SC
fixed point may exist else but the multiplicity of the conduction electron
channels. Recent works have shown that the presence of a more general Kondo
impurity where the spin symmetry is extended from $SU(2)$ to $SU(N)$, and
the representation is given by a L-shaped Young tableau, may also lead to an
instability of the SC fixed point \cite{cpt1,cpt2} already in the one-channel case. In the
large $N$ limit, Coleman {\it et al.} have found that the SC fixed point 
becomes
unstable as soon as $q$ (the number of boxes in the Young tableau along the
first column), is larger than $N/2$ whatever the value of $2S$  (the number
of boxes in the Young tableau along the first row) is. This result opens the 
route to the existence of an
intermediate coupling fixed point with presumably non-Fermi-liquid
excitation spectrum. The consideration of a L-shaped Kondo impurity fits in
with the supersymmetry approach that we have evoked before since both spin
operators and states can be expressed in terms of bosons and fermions.

At that point, it is worth noting that the supersymmetry theory, or
specifically taking into consideration more general L-shaped Kondo impurities
appears to offer valuable insights into the two issues raised by the
breakdown of the Fermi liquid theory that we have summarized above, i.e.
both the existence of locally critical modes and the question of the
instability of the SC fixed point. Somehow it seems that the choice
of L-shaped Kondo impurities captures the physics present in real systems,
with the coexistence of the screening of the spin by conduction electrons
responsible for the formation of quasiparticles, and the formation of a
localized magnetic moment that persists and eventually leads to a phase
transition as the coupling to other impurities becomes dominant. In the same
way as large N expansions may provide insights into real systems even at
finite value of the degeneracy, the study of more general impurities
may enlighten the understanding of experimental situations.
\bigskip 

The aim of the paper is to study the extended $SU(N)$ L-shaped
single-impurity Kondo model in the one-channel case.\ 
We want to understand how the system behaves, not only as a function of
the impurity parameters, $(2S,q)$, but also as a function of the 
number of electrons, $n_d$, available on neighboring sites,
that is to say, of the filling.
We find that as long as
the bosonic component of spin is of order $N$, there is a transition 
around the point where the fermionic 
component of the impurity is $q=N/2$. At this particular point, the energy 
shift
is, to lowest order in perturbation theory around the
strong-coupling fixed point, equal to $(-2t^2/J)$, independently of the
impurity parameters, $q$, $S$ and $N$. When $q<N/2$, the low-energy 
physics corresponds to a stable strong-coupling fixed point. For $q>N/2$
the strong-coupling fixed point is unstable and anomalous behavior is
expected, in particular, a two-stage quenching effect, as predicted by
Ref. \onlinecite{cpt1}. This phase
diagram is not accidental, but is due to the relation of the 
effective dressed impurity in the strong-coupling regime to the conduction
electrons in neighboring sites, as our study of the dependence of the
energy shifts on
$n_d$ reveals. If $q<N/2$, the energy is minimized when the dressed
impurity repels the electrons on the next site. That is,
when $n_d=1$. At $q=N/2$, the energy shift is independent of $n_d$.
Finally, the lowest energy shift for $q>N/2$ corresponds to a maximal
$n_d$, indicating the accumulation of conduction electrons on neighboring
sites, leading to a two-stage Kondo quenching.
\smallskip

The rest of the paper is organized as follows. 
In Section \ref{model}, we introduce
the model and the main features of the strong-coupling limit, where the
electron kinetic term is neglected. In this limit the model is reduced to
a single site problem, where the impurity is coupled to $n_c$ conduction
electrons. We identify the ground state and the energies of the excited 
states with one more or one less conduction electron, which will play a 
role in the lowest order in
perturbation theory. In section \ref{stability}, 
we discuss the stability of
the strong coupling fixed point as resulting from the sign of the effective
coupling $J_{ef\!f}$ between the spin of the strong coupling composite and
the spin of the conduction electrons on the neighboring site. The
calculation is based on a second order perturbation theory in $t$ and is
performed for an arbitrary number $n_{d}$ of conduction electrons on the
neighboring site. 
Section \ref{dissc} contains the discussion of the results.
In the large N limit, we show how $J_{ef\!f}$ is derived
from the energy shift difference between the symmetric and the antisymmetric 
configurations,
and how the analysis of the $n_{d}$ dependence of the energy shift provides
information on the nature of Kondo screening, with the realization of a
two-stage Kondo effect when the SC fixed point becomes unstable.\ When the
behavior of the system is controlled by the strong coupling fixed point, i.e.
when $q<N/2$, the impurity in the ground state tends to \textit{repel}
electrons on neighboring sites. Once $q>N/2$, the repulsion becomes
attraction. We show how this feature is already present in the purely
fermionic case, and is a consequence of the \textit{particle-hole} symmetry
(more details are contained in Appendix \ref{fermi}).
The fact that there is extra degeneracy in the supersymmetric impurity, due
to the bosonic contribution, leads to the instability of the strong coupling
fixed point as soon as $q>N/2$.\ We finish the section with a short
discussion on the behavior of physical quantities in the
different regimes.

The appendices contain the technical details of the calculations. In 
Appendix \ref{su3} we outline the construction of three particle states
with $SU(3)$ symmetry, as an introduction to the group theoretical 
formalism used. Explicit
expressions for the impurity states and the eigenstates of the model
in the strong coupling limit are derived in Appendix \ref{apu}. 
We  also include a general presentation of the different representations of
the spin, either bosonic, fermionic or L-shaped, as considered in the paper.
We will show how in the latter case the spin operators and the impurity
states are expressed in terms of fermion and boson creation and annihilation
operators within two constraints.
Appendix \ref{fermi} contains
a calculation of the energy shift to the strong coupling fixed point to
lowest order in perturbation theory, for the completely antisymmetric
impurity. Since the ground state is a singlet, there is no splitting of
levels. Nevertheless, the behavior of the energy with the filling, $n_d$,
on the neighboring site shares many common features with the problem that
we have studied. Finally, we include the details of the calculation of 
the matrix elements needed in the second order perturbation theory
calculation in Appendix \ref{matrix}.

\section{The model and its strong-coupling limit}
\label{model}

\subsection{SU(N) single-impurity Kondo model}

We consider a generalized, single-impurity, Kondo model with one channel of
conduction electrons and a spin symmetry group extended from $SU(2)$ to 
$SU(N)$. An impurity spin, $\mathbf{S}$, is placed at the origin (site $0$).
In this article we will deal with impurities that can be
realized by a combination of bosonic and fermionic operators, and are thus
described by a L-shaped representation in the language of Young tableaux \cite{hamm,jon,corn}, as
illustrated in Fig. \ref{yt1} (for details, see Appendix \ref{apu}).
\begin{figure}[h]
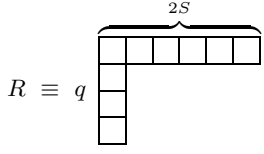

\begin{eqnarray*}
R~\equiv~q~\overbrace{\yng(6,1,1,1)}^{2S} 
\end{eqnarray*}
\caption{ 
\label{yt1} Young Tableau description of an impurity with mixed symmetry, 
$[2S,1^{q-1}]$, realized by a combination of fermions and bosons}
\end{figure}

If $2S$ and $q$ are the numbers of boxes along the first row and the first
column, respectively, the representation is denoted by $[2S,1^{q-1}]$. Its
degeneracy \cite{lich} is reported in Table 1. The
conduction electrons transform under the fundamental representation of 
$SU(N) $ and can be represented by Young tableaux made out of single boxes. 
The dimension of the fundamental representation is $N$ which just means that
each electron can be in one of $N$ states of spin.

The hamiltonian describing the model is written as 
\begin{equation}
H=\sum_{\mathbf{k},\alpha }\varepsilon _{\mathbf{k}}c_{\mathbf{k},\alpha
}^{\dagger }c_{\mathbf{k},\alpha }+J\sum_{A}\mathbf{S}^{A}\sum_{\alpha
,\beta }c_{\alpha }^{\dagger }(0)\boldsymbol {\tau}_{\alpha \beta
}^{A}c_{\beta }(0)~,  \label{ham1}
\end{equation}
where $c_{\mathbf{k},\alpha }^{\dagger }$ is the creation operator of a
conduction electron with momentum $\mathbf{k}$, and $SU(N)$ spin index 
$\alpha =a,b,...,r_N$, $c_{\alpha }^{\dagger }(0)=\frac{1}{\sqrt{N_{S}}}
\sum_{\mathbf{k}}c_{\mathbf{k},\alpha }^{\dagger }$ is the creation 
operator of a
conduction electron at the origin, $N_{S}$ is the number of sites, and 
$\boldsymbol{\tau }_{\alpha \beta}^{A}$ ($A=1,\dots,N^{2}-1$) are the 
generators of the $SU(N)$ group in the fundamental representation,
with $Tr[\boldsymbol{\tau}^{A}\boldsymbol{\tau}^{B}]=\delta _{AB}/2$. 
In the $SU(2)$ case, $\boldsymbol{\tau}^{A}=\boldsymbol{\sigma}^{A}/2$,
where $\{\boldsymbol{\sigma}^{A}\}$ are the Pauli matrices. The
conduction electrons interact with the impurity spin $\mathbf{S}^{A}$ 
($A=1,\dots,N^{2}-1$), placed at the origin, via Kondo coupling, $J>0$. 
When the
impurity is in the fundamental representation, we recover the
Coqblin-Schrieffer model \cite{csm} \cite{hew} describing conduction electrons 
in interaction with
an impurity spin of angular momentum $j$, ($N=2j+1$), resulting of the
combined spin and orbit exchange scattering. In our notation, $a=j$,
$b=j-1$,..., $r_N=-j$.

\subsection{Strong-Coupling fixed point}
\label{sscfp}

\begin{figure}[h]
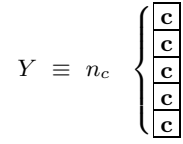

\begin{eqnarray*}
Y ~\equiv~ n_{c}~~ \left\{ 
\young(\bfc,\bfc,\bfc,\bfc,\bfc) \right.
\end{eqnarray*}
\caption {\label{ync}
Young tableau description of $n_c$ conduction electrons, {\it localized}
at the impurity site}
\end{figure}

In the strong-coupling limit, the hamiltonian reduces to the local Kondo
interaction term at site 0 
\begin{equation}
H=J\sum_{A}\mathbf{S}^{A}\sum_{\alpha ,\beta }c_{\alpha }^{\dagger }(0)%
\boldsymbol{\tau}_{\alpha \beta }^{A}c_{\beta }(0)~,
\end{equation}%
The ground state, $|GS\rangle $, is formed by binding the right amount of
conduction electrons to the impurity in order to minimize the Kondo energy. Let us
denote by $Y$ (Fig. \ref{ync}) the representation of the $n_{c}$ conduction 
electrons coupled to the impurity, $R$ that of the free 
impurity (Fig. \ref{yt1}), 
and $R_{SC}$ the representation of
one of the strong-coupling states resulting of the direct product $R \otimes
Y $ (cf. Fig. \ref{rs}) (see Appendix \ref{apu} for details).

\begin{figure}[h]
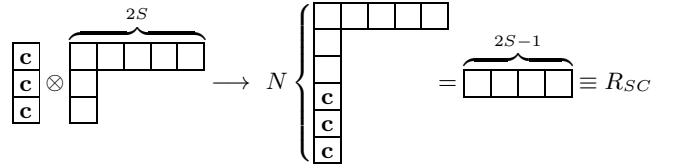

\begin{eqnarray*}
\young(\bfc,\bfc,\bfc)  
\otimes \overbrace{\yng(5,1,1)}^{2S}\longrightarrow~ N\left\{  
\young(~~~~~,~,~,\bfc,\bfc,\bfc) \right. \!\!\!\!\! = 
\overbrace{\yng(4)}^{2S-1} \equiv R_{SC}
\end{eqnarray*}
\caption {\label{rs}
Young tableau description of the formation of the strong coupling
ground state. We denote the presence of conduction electrons at site 0 
by $\bfc$. Notice that the first column in the Young 
tableau for $R_{SC}$ is a singlet and can be removed.}
\end{figure}
When $N=2$, the Kondo energy can be written in terms of
conserved quantities 
\begin{eqnarray*}
\lefteqn{J~\vec{\mathbf{S}}\cdot \sum_{\alpha ,\beta }c_{\alpha }^{\dagger }(0)\vec{%
\boldsymbol{\tau}}_{\alpha \beta }c_{\beta }(0)~|GS\rangle=} && \\
&&\frac{J}{2}\left[
S^{SC}(S^{SC}+1)-S^{R}(S^{R}+1)-S^{Y}(S^{Y}+1)\right] |GS\rangle~,
\end{eqnarray*}%
where $S(S+1)$ is the eigenvalue of the Casimir operator $\widehat{S}^{2}$
for $N=2$. The generalization to $SU(N)$ is given by 
\begin{eqnarray}
\lefteqn{J\sum_{A}\mathbf{S}^{A}\sum_{\alpha ,\beta }c_{\alpha }^{\dagger }(0)
\boldsymbol{\tau}_{\alpha \beta }^{A}c_{\beta }(0)~|GS\rangle =} &&
\nonumber \\ &&
\frac{J}{2}
\left[ \hat{\mathcal{C}_{2}}(R_{SC})-\hat{\mathcal{C}_{2}}(R)-\hat{\mathcal{C
}_{2}}(Y)\right] ~|GS\rangle~,
\label{kcas}
\end{eqnarray}
where $\mathcal{C}_{2}(\hat {R})$ is the quadratic Casimir operator of the
representation $\hat {R}$, defined earlier, which commutes with all the generators
of the group. For a representation given by a Young Tableau with $m_{j}$
boxes in the $j$-th row until the row $j=h$, the eigenvalue 
$\mathcal{C}_{2}(\{m_j\})$ of the quadratic Casimir operator is 
\begin{equation*}
\mathcal{C}_{2}(\{m_j\})=\frac{1}{2}\left[ \frac{Q(N^{2}-Q)}{N}%
+\sum_{j=1}^{h}m_{j}(m_{j}+1-2j)\right]
\end{equation*}
where $Q=\sum_{j=1}^{h}m_{j}$ is the total number of boxes \cite{oku}. 
Table \ref{casi}  summarizes the expressions of the Casimir eigenvalues 
for the impurities described in this work and for the conduction electrons, as well 
as the dimension of their spin representations.
\begin{table*}[h]
\caption{\label{casi} Dimension, $d$, and eigenvalues of the casimir operator,
$\mathcal{C}_{2}$ for the symmetric, antisymmetic,  L-shaped, and
fundamental representations
studied in this paper. In the L-shaped case, $Q=(2S+q-1)$ is the 
total number of boxes in the Young tableau, and $Y'=(q-2S)$ measures 
the row-column asymmetry} 
\begin{ruledtabular}
\begin{tabular}{l|cccc}  &&&& \\
& Symmetric & Antisymmetric & L-shaped & Fundamental \\ & $[2S]$ & $[1^{q}]$
& $[2S,1^{q-1}]$ & $[1]$ \\ &&&& \\  \hline &&&& \\ 
$~~d~~$ & $C_{N+2S-1}^{2S}$ & $C_{N}^{q}$ & $\left(\frac{2S}{2S+q-1}\right)
C_{N+2S-1}^{2S}C_{N-1}^{q-1}$ & $N$ \\ &&&& \\
$~~\mathcal{C}_{2}~~$ & $\frac{1}{2N}(2S(2S+N)(N-1))$ 
& $\frac{1}{2N}(q(N-q)(N+1))$  & 
$\frac{Q}{2}\left( N-Y^{\prime}-Q/N\right)$ & $\frac{1}{2N}(N^2-1)$ \\ &&&& \\
\end{tabular} 
\end{ruledtabular}
\end{table*}

Minimization of the 
energy Eq. (\ref{kcas}),
leads to a ground state with $n_{c}=(N-q)$ conduction electrons coupled to
the L-shaped Kondo impurity ensuring partial screening. 
The resulting composite at site 0, with energy $E_0$,
is made out of the impurity dressed by the conduction electrons in order to
form a singlet along the first column. The associated Young tableau in the
strong coupling regime is given in Fig. \ref{rs}.
Note that the first column of length $N$ can be removed without changing the
representation since it is a singlet. 
When the strong-coupling fixed point
is stable, this corresponds to a one-stage Kondo effect in which the
impurity is screened by the conduction electrons to form a bosonic $(S-1/2)$
impurity. 

\subsection{Ground state}
\label{subgs}

Let us now write the expression of the fundamental state associated with this
strong-coupling fixed point. The ground state is degenerate. The states in
the multiplet transform as a completely symmetric representation of $SU(N)$,
described by a Young tableau with $(2S-1)$ boxes, denoted by $[2S-1]$,
(Fig. \ref{rs}).
We choose a realization of the impurity in terms of $2S$ bosonic operators and
$(q-1)$ fermionic operators, which happens to be more convenient. We could have
constructed 
impurity states with the same $SU(N)$ symmetry using $(2S-1)$ bosons and $q$ 
fermions (see Appendix \ref{apu}).
We would like to emphasize that all the results that we establish in this 
paper 
are independent of the operator representation which we choose to work with.
The highest weight state is then written as 
\begin{equation}
|GS\rangle _{\{a\}aa}^{[2S-1]}=\frac{1}{\scriptstyle\sqrt{(2S-1)!}}%
(b_{a}^{\dagger })^{2S-1}|\Delta \rangle~ , \label{grst}
\end{equation}%
with 
\begin{equation}
|\Delta \rangle \equiv \frac{1}{\gamma }\mathcal{A}(b_{i_{1}}^{\dagger
}(\prod_{\alpha =i_{2}}^{i_{q}}f_{\alpha }^{\dagger })(\prod_{\beta
=i_{q+1}}^{i_{N}}c_{\beta }^{\dagger }))|0\rangle~ ,
\label{delta}
\end{equation}
\begin{equation*}
\gamma \equiv \sqrt{%
(2S+N-1)C_{N-1}^{q-1}}~.
\end{equation*}
Here, $|\Delta \rangle $
transforms itself as a $SU(N)$ singlet and it will be annihilated
by any of the raising and lowering operators, $T^{\pm }|\Delta \rangle
=U^{\pm }|\Delta \rangle =\cdots =0$. This ``state'' would describe the
strong coupling ground state for a purely fermionic impurity. 

\subsection{Excited States}
\label{exst}

There are two types of excited states in the strong-coupling regime. Either
the degenerate ground state acquires an additional conduction electron at
the impurity site, $|GS+1\rangle $, or it
loses one conduction electron, 
$|GS-1\rangle $. In the former case, $|GS+1\rangle $, the spin of the 
additional conduction electron can be either symmetrically or 
antisymmetrically correlated with
the spin of the impurity as schematized in Fig. \ref{fx}. 
\begin{figure}[h]
\begin{equation*}
|GS+1\rangle ^{S}~\equiv~\young(~~~~\bfc,~,~,\bfc,\bfc,\bfc)
~=~~\young(~~~\bfc) 
\end{equation*}
\begin{equation*}
|GS+1\rangle ^{A}~\equiv~~\young(~~~~,~\bfc,~,\bfc,\bfc,\bfc)
~=~~\young(~~~,\bfc)
\end{equation*}
\caption {Excited states, $|GS+1\rangle^S$, $|GS+1\rangle^A$, with an
additional conduction electron, $n_c=(N-q+1)$, respectively in the 
symmetric and
antisymmetric configurations.} 
\label{fx}
\end{figure}

In the limiting case of $SU(2)$ spin, these two configurations correspond to
a spin of the conduction electron that is either parallel or antiparallel to
the impurity spin. In the general $SU(N)$ case, we will keep on speaking of
symmetric and antisymmetric configurations respectively. 

States with one less electron will be denoted by $|GS-1\rangle$ and are
represented by the Young tableau in Fig. \ref{fy}.
\begin{figure}[h]
\begin{equation*}
|GS-1\rangle ~\equiv~ (N-1)\left\{ \young(~~~~~,~,~,\bfc,\bfc) \right. \nonumber
\end{equation*}
\caption {Excited state, $|GS-1\rangle$, with one less conduction electron 
$n_c=(N-q-1)$.}
\label{fy}
\end{figure}
Let us denote by $\Delta E_{1}^{S}= E_{1}^{S}-E_0$, $\Delta E_{1}^{A}= E_{1}^{A}-E_0$ and 
$\Delta E_{2}=E_{2}-E_0$ the energy differences, with respect to the ground state
energy,
associated with these three excited states $|GS+1\rangle ^{S}$, 
$|GS+1\rangle ^{A}$ and $|GS-1\rangle$. Using the same casimirology method 
as presented at the beginning of this section 
for the determination of the ground state energy, we have summarized our 
results in Table \ref{tde} respectively for arbitrary N and in the 
large-N limit with $(2S+q-1)/N$ finite.

\begingroup
\squeezetable
\begin{table}[h]
\caption {Strong coupling
excitation energies, $\Delta E^S_1$,  $\Delta E^A_1$, and $\Delta E_2$,
in the case of an L-shaped impurity,
measured with respect to the ground state, of the
states with one more
conduction electrons on site 0, coupled symmetrically and antisymmetrically,
respectively, to the dressed impurity on site 0, and of the state 
with one less electron. }
\label{tde}
\begin{ruledtabular}
\begin{tabular}{c|ccc} &&& \\
& $\Delta E_1^S$ & $\Delta E_1^A$ & $\Delta E_2$ 
\\ &&& \\  \hline &&& \\ 
Arbitrary N & 
$\frac{J}{2}(2S+N-q-Q/N)$ & 
$\frac{J}{2}(N-q-Q/N)$ &
$\frac{J}{2}(q+Q/N)$
\\ &&& \\
Large N limit \\
($Q/N$ finite) & $\frac{J}{2}(N-q+2S)$ 
& $\frac{J}{2}(N-q)$  & 
$\frac{J}{2}q$ \\ &&& \\
\end{tabular} 
\end{ruledtabular}
\end{table}
\endgroup

One can check that the results in Table \ref{tde} coincide with 
(Eqs.[25-26]) in ref. \onlinecite{cpt2}, within a $N/2$ factor
stemming from a different definition of the Kondo coupling $J$ (cf. Eq.(1)
of ref. \onlinecite{cpt2}) and of the Casimir (cf. Eq.(17) of ref. 
\onlinecite{cpt2}), 
and a change in the
notations $n_{f}^{\ast }=q$ and $n_{b}=2S$.

\section{Stability of the strong coupling fixed point}
\label{stability}

We have identified the strong coupling fixed point in the previous
section. 
For $J\rightarrow \infty$, the lowest energy state corresponds
to $n_c=(N-q)$ electrons partially screening the impurity at the origin,
and free electrons in the other sites, unable to hop to the impurity site.

In order to better understand the low-energy physics of the system,
we should consider the finite Kondo coupling, allowing virtual hopping from
and to the impurity site. These processes generate interactions between the
composite at site 0 and the conduction electrons on neighboring sites, that 
can be treated as perturbations of the strong coupling fixed point. Applying 
an analysis similar to that of Nozi\`eres and Blandin \cite{nb}
to the nature of the 
excitations, we can argue whether or not the strong coupling fixed point 
remains stable once virtual hopping is allowed. 

\begin{figure}[h]
\includegraphics[width=8.0cm]{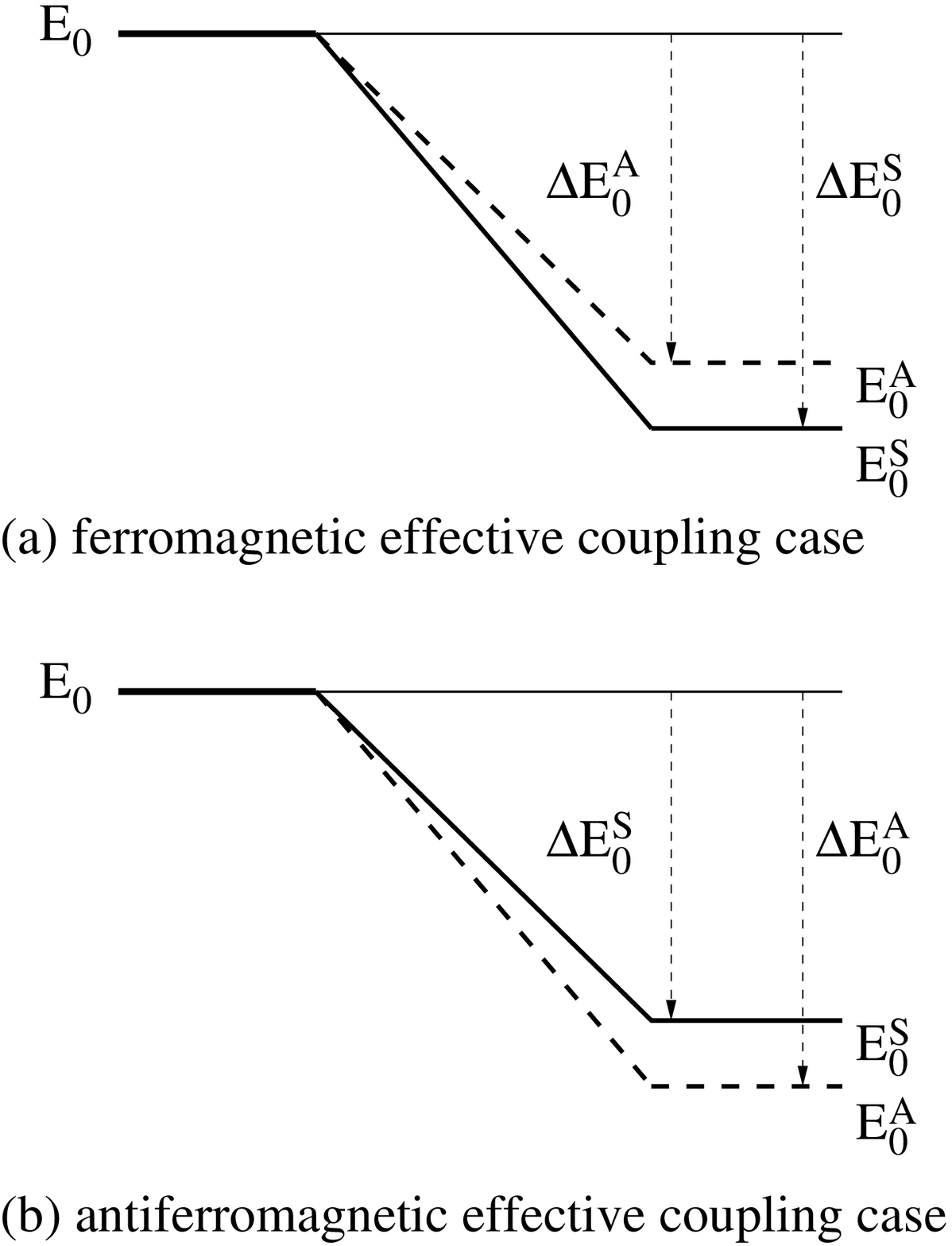}
\caption {Second order perturbation theory energy shift of the strong coupling ground state 
in the cases of a ferromagnetic (Fig.a) and antiferromagnetic (Fig.b)
effective coupling}
\label{ener2}
\end{figure}

We consider a system with an additional site next to the dressed impurity, 
filled with $n_d$ electrons. 
The ground state consists of two multiplets, with different symmetry
properties. Once the hopping is turned on, the degeneracy is
lifted, and each multiplet acquires a different energy shift denoted by
$\Delta E^S$ and $\Delta E^A$, respectively (see Fig. \ref{ener2}).
We can reproduce this spectrum by considering an effective coupling 
between the spin of the dressed
spin at site-0, and the spin of the $n_d$ electrons on site 1. If $E^S$ lies
(above)below $E^A$, the effective coupling is (anti)ferromagnetic. 

Thus, if the coupling 
between the effective spin at the impurity site and that of the electrons
on site 1 is ferromagnetic we know, from the scaling analysis at weak 
coupling, that the perturbation is irrelevant, and the low energy physics
is described by the strong coupling fixed point. That is, an underscreened, 
completely symmetric, effective
impurity weakly coupled to a gas of free electrons with a phase shift 
indicating that there are already $(N-q)$ electrons screening the original impurity.
In the completely antisymmetric case ($2S=1$), the phase shift corresponds to
the unitary limit, $\delta=\pi/2$, 
for $SU(2)$, and is a function \cite{pgks} of $q/N$  for $SU(N)$, 
reaching the unitary limit 
for $q=N/2$ (see appendix \ref{fermi}).

If, on the contrary, the effective coupling is antiferromagnetic, the perturbation
is relevant, the strong coupling fixed point is unstable and the 
low-energy physics of the model corresponds to some intermediate coupling fixed 
point,
to be identified. The authors in Refs. \onlinecite{cpt1} and 
\onlinecite{cpt2}, 
have claimed that the flow away from
the strong coupling fixed point is characterized by a {\it two-stage
quenching} of the impurity. After the impurity has been partially 
screened by the $(N-q)$ conduction electrons on site 0, $(N-1)$ additional 
conduction electrons tend
to accumulate around the impurity, leading to further screening.

In this section we explicitely calculate the effects of hopping on the
strong coupling fixed point to the lowest order in perturbation theory,
that is, second order in $t$. We will consider
the case with an arbitrary number $n_d$ of conduction electrons in site 1
generalizing the case $n_d=1$ considered in Ref. \onlinecite{cpt2}. This 
will allow us to understand
the origin of the instability of the strong coupling fixed point and 
eventually enlighten the question about the nature of the Kondo screening :
either one-stage or two-stage Kondo effect depending on which regime is 
considered.
\begin{figure}[h]
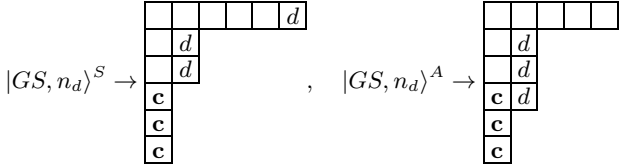

\begin{eqnarray*}
|GS,n_d\rangle^S \rightarrow \young(~~~~~d,~d,~d,\bfc,\bfc,\bfc),
~~~
|GS,n_d\rangle^A \rightarrow \young(~~~~~,~d,~d,\bfc d,\bfc,\bfc)
\end{eqnarray*}
\caption {Strong coupling ground state in the presence of $n_d$ conduction electrons
on site 1 coupled either symmetrically or antisymmetrically to the dressed impurity
on site 0}
\label{sa}
\end{figure}

Before turning on the hopping term, let us consider {\it ground states} of the
form $|GS,n_d\rangle=\sum |GS\rangle_0|n_d\rangle_1$, with $n_d$ electrons on 
site 1. According to the $SU(N)$ symmetry there are two possible 
configurations, depending on whether the $n_d$ electrons are coupled
symmetrically or antisymmetrically to the composite on site 0. This 
corresponds to the Clebsch-Gordan series $[2S-1]\otimes[n_d] \rightarrow
[2S,1^{n_d-1}]\oplus [2S-1,1^{n_d}]$. We denote the states by
$|GS,n_d\rangle^S$, and $|GS,n_d\rangle^A$, respectively (Fig. \ref{sa}).

The $SU(N)$ symmetry is preserved by the hopping. That means that the
perturbation will shift the energies of $|GS,n_d\rangle^S$ and 
$|GS,n_d\rangle^A$ separately, without mixing the states. We will thus 
denote the 
shifts by $\Delta E_0^S$ and $\Delta E_0^A$, respectively.

The hopping term is of the form
\begin{eqnarray*}
H_h = H_1+H_2 = t\sum_\alpha c^\dagger_\alpha d_\alpha + t 
\sum_\alpha d^\dagger_\alpha c_\alpha,~~
(H_1)^\dagger = H_2
\end{eqnarray*}
where $d^\dagger_\alpha$ creates an electron on site 1. We can distinguish two
types of processes, corresponding to different intermediate states. The
first type, which we denote process 1, corresponds to an electron hopping
from site 1 into site 0 first, probing excited states $|GS+1\rangle^{S,A}$,
and then hopping back to site 1. The indices $S,A$ correspond to the two
possible intermediate states depending whether the conduction electron
which hops to site 0 is symmetrically or antisymmetrically correlated with 
the dressed impurity as we will see in details in next section. The 
contribution of the process 1 to the energy
shift is the following
\begin{widetext}
\begin{eqnarray*}
t^2 \sum_{\alpha,\beta}\sum_i 
\frac{\langle GS,n_d|d^\dagger_\beta c_\beta|GS+1,n_d-1\rangle ^i \,
{}^i\langle GS+1,n_d-1| c^\dagger_\alpha d_\alpha|GS,n_d\rangle}
{(E_0-E_1^i)}~,~
\end{eqnarray*}
with $i=S,A$ (see Appendix \ref{apu}). 

In process 2, the electron hops from 
site 0 to site 1 first and then back to site 0, 
probing $|GS-1\rangle$, leading to a
contribution to the energy shift of the form
\begin{eqnarray*}
t^2 \sum_{\alpha,\beta} 
\frac{\langle GS,n_d|c^\dagger_\beta d_\beta|GS-1,n_d+1\rangle
\langle GS-1,n_d+1| d^\dagger_\alpha c_\alpha|GS,n_d\rangle}
{(E_0-E_2)}~,~
\end{eqnarray*}
\end{widetext}
Hence, the energy shifts for the symmetric and antisymmetric configurations 
are, respectively,
\begin{equation}
\Delta E_{0}^{S}=\frac{M_{1}^{S}}{E_{0}-E_{1}^{S}}+\frac{\overline{M_{1}^{S}%
}}{E_{0}-E_{1}^{A}}+\frac{M_{2}^{S}}{E_{0}-E_{2}}, 
\label{pert1}
\end{equation}
\begin{equation}
\Delta E_{0}^{A}=\frac{M_{1}^{A}}{E_{0}-E_{1}^{A}}+\frac{M_{2}^{A}}
{E_{0}-E_{2}},
\label{pert2} 
\end{equation}
where the expressions in the denominators, $(E_{0}-E_{1}^{S})=%
-\Delta E_1 ^S$,  
$(E_{0}-E_{1}^{A})=%
-\Delta E_1 ^A$, and
$(E_{0}-E_{2})=%
-\Delta E_2$ measuring the energy of the excited states compared 
to the energy of the ground state are given in Table \ref{tde}. The matrix elements, $M$,
will be introduced below as we will study the contribution of each process.
The energy difference between the two states,
\begin{eqnarray}
\Delta E_0^S-\Delta E_0^A = \frac{M_{1}^{S}}{E_{0}-E_{1}^{S}}
+\frac{\overline{M_{1}^{S}}-M_1^A}{E_{0}-E_{1}^{A}}
+\frac{M_{2}^{S}-M_2^A}{E_{0}-E_{2}}~, \nonumber \\
\label{fase1}
\end{eqnarray}
determines the sign of the effective interaction and the stability of the
strong coupling fixed point.

\subsection{Process 1, symmetric configuration}

We consider first the case where the $n_d$ electrons in the site-1 are
coupled to the site-0 state in the most symmetric configuration. In the
shorthand notation that we use for the Young Tableaux, it corresponds to the
state $[2S-1]\otimes[1^{n_d}]\rightarrow[2S,1^{n_d-1}]$. Here, as opposed to
the case $n_d=1$ \cite{cpt2}, the hamiltonian transforms the ground state 
\begin{eqnarray}
|GS,n_d\rangle^{S} = (d^\dagger_a d^\dagger_b \cdots d^\dagger_u)
|GS\rangle~,  \label{gsd}
\end{eqnarray}
into a linear combination of two excited states : 
$|GS+1,n_d-1\rangle^{S}$, with energy $E_S$, and 
$|\overline{GS+1,n_d-1}
\rangle^{S}$, with energy $E_A$ depending on whether the additional
conduction electron in site 0 is coupled symmetrically or antisymmetrically to
the dressed impurity. 
\begin{figure}
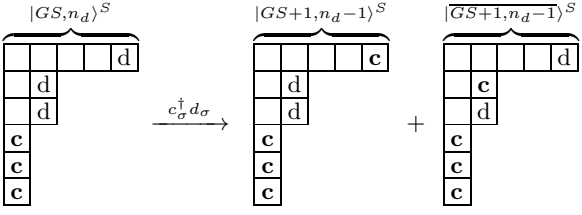

\begin{eqnarray*}
\begin{CD} \overbrace{\young(~~~~\rmd,~\rmd,~\rmd,\bfc,\bfc,\bfc)}^
{|GS,n_d\rangle^S}  @>c^\dagger_\sigma d_\sigma>> ~
\overbrace{\young(~~~~\bfc,~\rmd,~\rmd,\bfc,\bfc,\bfc)}
^{|GS+1,n_d-1\rangle^{S}} ~+~
\overbrace{\young(~~~~\rmd,~\bfc,~\rmd,\bfc,\bfc,\bfc)}^
{|\overline{GS+1,n_d-1}\rangle^{S}} \end{CD}
\end{eqnarray*}
\caption {When $n_d$ conduction electrons are coupled symmetrically to the
dressed impurity at the origin, the term $c^\dagger_\sigma 
d_\sigma$, acting on the ground state, 
generates a linear combination of two excited states with an
additional conduction electron at the origin. }
\label{p1s}
\end{figure}
The state obtained by acting with $c^\dagger_\sigma d_\sigma$ on 
the ground state defined by Eq. \ref{gsd}
has to be computed explicitely, and the result written as a linear
combination of the excited states, Fig. \ref{p1s}. The latter are 
obtained by coupling the
site-0 states with $(n_d-1)$ conduction electrons on site 1. 
The explicit expressions for
these states are given in Appendix \ref{matrix}, and they lead to the result 
\begin{eqnarray}
\lefteqn{\left(\sum_\sigma c^\dagger_\sigma d_\sigma\right)
|GS,n_d\rangle^{S}} && \nonumber \\  &=& \Omega \sqrt{\frac{2S+n_d-1}{2S}}%
~|GS+1,n_d-1\rangle^{S}  \nonumber \\
&+& \Lambda \sqrt{n_d-1} \sqrt{\frac{2S-1}{2S}}~ 
|\overline{GS+1,n_d-1}
\rangle^{S}~.  \label{symm}
\end{eqnarray}
Where the normalization coefficients \cite{cpt2}, 
$\Omega=\sqrt{\frac{2S+q-1}{2S+N-1}}$ and $\Lambda = \sqrt{\frac{q-1}{N-1}}$ 
are
independent of $n_d$, as we are considering hopping 
of a single electron. From here, we obtain the 
following matrix elements: 
\begin{eqnarray}
M^S_1 &=& |{}^S\langle GS+1,n_d-1|H_1|GS,n_d\rangle^S|^2 \nonumber \\
&=& t^2\left(\frac{2S+n_d-1}{2S}\right)\left(\frac{2S+q-1}{2S+N-1}\right)~,~
\label{m0a}
\end{eqnarray}
\begin{eqnarray}
\overline{M^S_1} &=& |{}^S\langle \overline{GS+1,n_d-1}|H_1|GS,n_d\rangle^S|^2
\nonumber \\
&=& t^2(n_d-1)\left(\frac{2S-1}{2S}\right) 
\left(\frac{q-1}{N-1}\right)~.~  \label{m1a}
\end{eqnarray}
We see right away that $\overline{M}^S_1$ is proportional to $(n_d-1)$, and
vanishes for $n_d=1$, whereas $M^S_1$ depends noticeably on $n_d$ only for $%
2S \ll n_d < N$.

\subsection{Process 1, antisymmetric configuration}

\begin{figure}
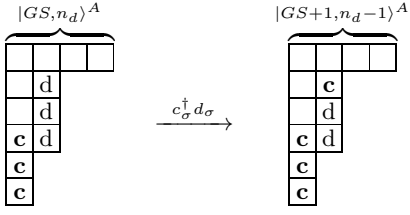

\begin{eqnarray*}
\begin{CD} \overbrace{\young(~~~~,~\rmd,~\rmd,\bfc\rmd,\bfc,\bfc)}
^{|GS,n_d\rangle^{A}} ~~~@>c^\dagger_\sigma d_\sigma>>~~~
\overbrace{\young(~~~~,~\bfc,~\rmd,\bfc\rmd,\bfc,\bfc)}
^{|GS+1,n_d-1\rangle^{A}}~. \end{CD}
\end{eqnarray*}
\caption {When $n_d$ conduction electrons are coupled antisymmetrically to the
dressed impurity at the origin, the term $c^\dagger_\sigma 
d_\sigma$, acting on the ground state, 
generates state proportional to a given excited state, with
additional conduction electron at the origin. }
\label{p1a}
\end{figure}

Next, we consider the case where the electrons on site 1 are coupled to the
effective spin on site 0 according to $[2S-1]\otimes[1^{n_d}] \rightarrow
[2S-1,1^{n_d}]$. In the previous section it was easy to write down the
strong coupling ground state by just putting together the effective spin and
the $n_d$ electrons in the highest weight state possible, to obtain Eq.
(\ref{gsd}). Here we have to work out the necessary Clebsch-Gordan coefficients.
We present some of these coefficients in Table \ref{bigt} of Appendix
\ref{matrix}. The explicit
form of the ground state is 
\begin{widetext}
\begin{equation}
|GS,n_d\rangle^{A}_{abc\dots v}
= \frac{1}{\sqrt{2S+n_d-1}}\left(\sqrt{2S-1} (\prod_{i=2}^{n_d+1}
d^\dagger_{y_i})|GS\rangle_{aa}+\sum_{j=2}^{n_d+1}(-1)^{j-1} ({\prod_{i=1,
i\ne j}^{n_d+1}} d^\dagger_{y_i})|GS\rangle_{ay_j}\right)~, \label{gsad}   
\end{equation}
\end{widetext}
in the notation $y_1=a$. As in the $n_d=1$ case, the hopping term transforms 
the state defined by Eq. \ref{gsad} into a
state proportional to a given strong-coupling excited state (Fig. \ref{p1a}), 
In order to obtain the corresponding matrix element, we have computed
explicitely 
\begin{equation*}
\left(\sum_\sigma c^\dagger_\sigma d_\sigma\right)
|GS,n_d\rangle^{A}~ \propto ~ |GS+1,n_d-1\rangle^{A}~,
\end{equation*}
and then we have normalized the resulting state. The details can be found in
Appendix \ref{matrix}. We have 
\begin{equation*}
\left(\sum_\sigma c^\dagger_\sigma d_\sigma\right)
|GS,n_d\rangle^{A}~ = ~ \Lambda~ \sqrt{n_d}~
|GS+1,n_d-1\rangle^{A}~,
\end{equation*}
and the matrix element 
\begin{eqnarray}
M^A_1 &=& |{}^A\langle GS+1,n_d-1|H_1 
|GS,n_d\rangle^{A}|^2 \nonumber \\
&=& t^2\, n_d\left(\frac{q-1}{N-1}\right)~.  \label{ma}
\end{eqnarray}
Notice the dependence on $n_d$, and the fact that the matrix element does not
depend on $2S$. Combining together $\overline{M}^S_1$
and $M^A_1$ as it appears in Eq. (\ref{fase1}), we have 
\begin{eqnarray}
\overline{M}^S_1-M^A_1 = - t^2 
\left(\frac{2S+n_d-1}{2S}\right) \left( \frac{q-1}{N-1} \right)~.
\label{mp0a}
\end{eqnarray}
This is a term with the same $n_d$ dependence as $M^S_1$ but with the
opposite sign.

\subsection{Process 2, the trick}
\label{trick}

\begin{figure}
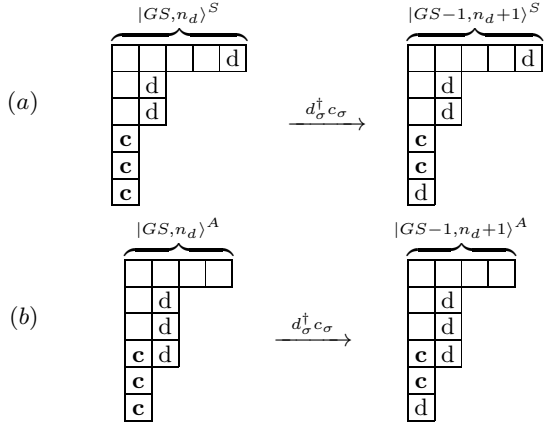

\begin{eqnarray*}
(a)~~~~~~~ & \begin{CD}
\overbrace{
\young(~~~~\rmd,~\rmd,~\rmd,\bfc,\bfc,\bfc)}^{|GS,n_d\rangle^{S}}~~~
@>d^\dagger_\sigma c_\sigma>> ~~~
\overbrace{\young(~~~~\rmd,~\rmd,~\rmd,\bfc,\bfc,\rmd)}
^{|GS-1,n_d+1\rangle^{S}}  \end{CD} \\ (b)~~~~~~~ & \begin{CD}
 \overbrace{\young(~~~~,~\rmd,~\rmd,\bfc\rmd,\bfc,\bfc)}
^{|GS,n_d\rangle^{A}} ~~~@>d^\dagger_\sigma c_\sigma>>~~~
\overbrace{\young(~~~~,~\rmd,~\rmd,\bfc\rmd,\bfc,\rmd)}
^{|GS-1,n_d+1\rangle^{A}} \end{CD}
\end{eqnarray*}
\caption {The term $d_\sigma^\dagger c_\sigma$, acting on the ground
state with $n_d$ electrons on site 1, coupled symmetrically, (a), or 
antisymmetrically, (b), to the impurity. The result is a state 
proportional to an excited state with one less electron at the
impurity site, coupled to $(n_d+1)$ electrons symmetrically, (a),
or antisymmetrically (b).}
\label{p2as}
\end{figure}

Both in the symmetric and antisymmetric configurations in
process 2 there is a one-to-one correspondence between the state obtained
by the action of $H_2$, and the excited state with the same symmetry
\cite{cpt2} (Fig. \ref{p2as}).
We write, for completeness, the Young Tableaux associated with these
processes
The evaluation of the two remaining matrix elements, $M_{2}^{S}$ and 
$M_{2}^{A}$, associated with process 2 is simplified by using the following
trick connecting the matrix elements of 
processes 1 and 2, respectively. On
the one hand, we have in the process 1
\begin{eqnarray*}
M_{1}^{S}+\overline{M_{1}^{S}} &=&t^{2}\sum_{\sigma \sigma ^{\prime
}}{}^{S}\!\langle GS,n_{d}| d_{\sigma ^{\prime }}^{\dagger
}c_{\sigma ^{\prime }}c_{\sigma }^{\dagger }d_{\sigma }|
GS,n_{d}\rangle^{S}   \\
&=&t^{2}\sum_{\sigma \sigma ^{\prime }} {}^S\!\langle d_{\sigma ^{\prime
}}^{\dagger }(\delta _{\sigma \sigma ^{\prime }}-c_{\sigma }^{\dagger
}c_{\sigma ^{\prime }})d_{\sigma }\rangle^S \\
&=&t^{2}\sum_{\sigma }{}^S\!\langle d_{\sigma }^{\dagger }d_{\sigma
}\rangle^S -\delta M^{S}~,
\end{eqnarray*}%
where $\delta M^{S}=t^{2}\sum_{\sigma \sigma ^{\prime }}
\left.\right.^S\!\langle
c_{\sigma }^{\dagger }d_{\sigma ^{\prime }}^{\dagger }d_{\sigma }c_{\sigma
^{\prime }}\rangle^{S} $ and the expectation value ${}^S\!\langle
{}\rangle^S $ should be considered over the ground state, $\left\vert
GS,n_{d}\right\rangle ^{S}$. On the other hand, the following property holds
for the matrix elements of process 2
\begin{eqnarray*}
M_{2}^{S} &=&t^{2}\sum_{\sigma \sigma ^{\prime }}{}^{S}\!\langle
GS,n_{d}| c_{\sigma ^{\prime }}^{\dagger }d_{\sigma ^{\prime
}}d_{\sigma }^{\dagger }c_{\sigma }| GS,n_{d}\rangle ^{S} \\
&=&t^{2}\sum_{\sigma \sigma ^{\prime }}{}^S\!\langle c_{\sigma ^{\prime
}}^{\dagger }(\delta _{\sigma \sigma ^{\prime }}-d_{\sigma }^{\dagger
}d_{\sigma ^{\prime }})c_{\sigma }\rangle^S \\
&=&t^{2}\sum_{\sigma }{}^S\langle c_{\sigma }^{\dagger }c_{\sigma
}\rangle^S -\delta M^{S}~,
\end{eqnarray*}%
where in the last line, we have exchanged the dummy variables $\sigma $ and $%
\sigma ^{\prime }$. The same relation exists for the antisymmetric
configuration (with expectation values taken on $|GS,n_d\rangle^{A}$)
\begin{eqnarray*}
M_{1}^{A} &=&t^{2}\sum_{\sigma }{}^A\!\langle d_{\sigma }^{\dagger }d_{\sigma
}\rangle^A -\delta M^{A}~, \\
M_{2}^{A} &=&t^{2}\sum_{\sigma }{}^A\!\langle c_{\sigma }^{\dagger }c_{\sigma
}\rangle^{A} -\delta M^{A}~.
\end{eqnarray*}
Thus, using the trick, we obtain
\begin{eqnarray}
M_2^S &=& t^2(n_c-n_d)+(M_1^S+\overline{M_1^S}) \nonumber  
\\ &=& t^2\,
 (N-n_d)\left( \frac{N-q}{N-1} \right) \left(\frac{2S+N-2}{2S+N-1}\right)~,~
\label{52}
\end{eqnarray}
\begin{eqnarray}
M_2^A &=& t^2(n_c-n_d)+M_1^A  \nonumber \\ &=& t^2\,
(N-(n_d+1)) \left(\frac{N-q}{N-1}\right)~, \label{53}
\end{eqnarray}
\begin{eqnarray}
M_2^S-M_2^A &=& M_1^S+\overline{M_1^S}-M_1^A \nonumber \\  &=&  t^2\,
\left(\frac{2S+n_d-1}{2S+N-1}\right)\left(\frac{N-q}{N-1}\right)~.~
\label{54}
\end{eqnarray}

\section{Discussion}
\label{dissc}

Once we have computed all the matrix elements, we can evaluate the
difference in energy shifts between the symmetric and the antisymmetric
configurations, $(\Delta E_0^S-\Delta E_0^A)$, induced by the perturbation
$H_h$, according to Equation (\ref{fase1}). We can then compare the result to an 
effective 
spin-spin interaction, with coupling $J_{ef\!f}$,
between the spin $\mathbf{S_0}$ of the dressed impurity at site 0, 
and the spin $\mathbf{S_1}$
of the $n_d$ electrons
at site 1. If the symmetric configuration is the lowest in energy (Fig.
\ref{ener2}(a)), it means
that the effective coupling is ferromagnetic and the perturbation is
irrelevant. Thus, the strong coupling fixed point is stable and allows to describe
the low-energy
behavior of the system. If, on the contrary, it is the antisymmetric
configuration the lowest in energy (Fig.
\ref{ener2}(b)), the effective coupling is
antiferromagnetic, and we know from scaling arguments that the perturbation
is relevant. The strong coupling fixed point is then unstable with respect
to hopping, and the behavior of the system is described by an intermediate
coupling fixed point. Incorporating the expressions of the matrix elements, 
Eqs.
(\ref{m0a},\ref{mp0a},\ref{54}), and 
those of the excitation energies (Table \ref{tde}) into 
Eq. (\ref{fase1}), 
one finds
\begin{widetext}
\begin{eqnarray}
\lefteqn{\Delta E_0^S-\Delta E_0^A = -(2S+n_d-1)\left(\frac{2t^2}{J}\right) } &&
\nonumber \\ && \times 
 \left\{ \frac{2S+q-1}{
2S(2S+N-1)(2S+N-q-(2S+q-1)/N)} \right. \nonumber \\ &&~~~
 + \frac{(N-q)}{(N-1)(2S+N-1)(q+(2S+q-1)/N)} 
- \left.\frac{q-1}{2S(N-1)(N-q-(2S+q-1)/N)}
\right\}~. \nonumber \\
\label{resu}
\end{eqnarray}
\end{widetext}
As the bosonic part increases, the energy difference becomes smaller. The dependence of 
the r.h.s. of Eq. (\ref{resu}) with $(2S+n_d-1)$ is linear and appears factored out. The effect of $n_d$ on 
$(\Delta E_0^S-\Delta E_0^A)$ is weak as long as $n_d\ll 2S$.
\begin{figure}[ht]
\includegraphics[width=8.0cm]{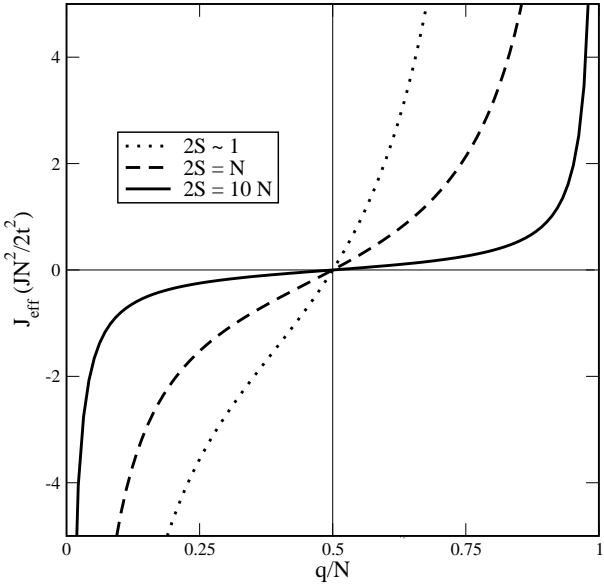}
\caption {Effective coupling, $J_{ef\!f}$, as a function of
$q/N$, for different values of $2S$, in the large-N limit.}
\label{fdel}
\end{figure}

We now calculate explicitely the interaction energy of two spins, in
the representations $[2S-1]$ and $[1^{n_d}]$, respectively, and identify the effective 
coupling $J_{ef\!f}$ from 
\begin{eqnarray*}
J_{ef\!f}\sum_A (\mathbf{S}_0^{[2S-1]})^A  (\mathbf{S}_1^{[1^{n_d}]})^A~,
\end{eqnarray*}
where $A=1,N^2-1$. The spectrum consists of two multiplets, according to 
the Clebsh-Gordan Series $[2S-1]\otimes[1^{n_d}]
\rightarrow [2S,1^{n_d-1}]^S\oplus [2S-1,1^{n_d}]^A$. We have added
superindices to indicate which is the {\it symmetric} state and
which is the {\it antisymmetric} one. As before, the energy is given
in terms of Casimir operators, Eq. (\ref{kcas}), and the energy difference
between the states is
\begin{eqnarray*}
\lefteqn{\Delta E_{[2S,1^{n_d-1}]}^S-\Delta E_{[2S-1,1^{n_d}]}^A} && 
 \\ &=&
\frac{J_{ef\!f}}{2} \left[ {\cal C}_2([2S,1^{n_d-1}])
-{\cal C}_2([2S-1,1^{n_d}])\right]  \\ &=& 
-\frac{J_{ef\!f}}{4}(2S+n_d-1)
\left[{Y'}_{[2S,1^{n_d-1}]}-{Y'}_{[2S-1,1^{n_d}]}\right]  \\
&=& 
\frac{J_{ef\!f}}{2}(2S+n_d-1)~,
\end{eqnarray*}
where we have used the results of Table \ref{casi}. Since both states have Young Tableaux with
the same number of boxes, $Q_{ef\!f}=2S+n_d-1$, the energy difference depends only on the
second constraint (\ref{qys}), $\hat{\cal{Y}}_{ef\!f}=Q_{ef\!f}Y'$
(see Appendix \ref{apu}). As a consequence,
the dependence on $n_d$ 
is factorized in the same way as in Eq. (\ref{resu}). Thus, we can identify
the effective coupling, $J_{ef\!f}$ with the term in curly brackets in
Eq. (\ref{resu}) times $4t^2/J$. In Fig. \ref{fdel} we plot this energy 
difference for several values of $S$.

\begin{figure}[ht]
\includegraphics[width=8.0cm]{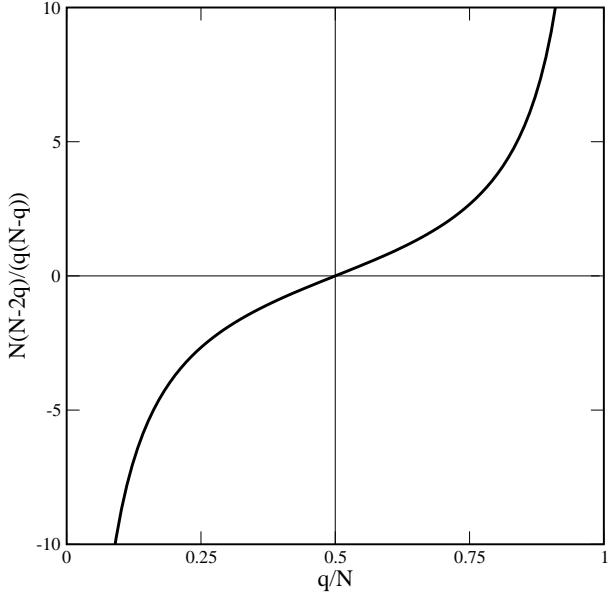}
\caption {Proportionality factor $\lambda=N(N-2q)/q(N-q)$ as a function of 
$q/N$. $\lambda$ determines both the sign of $J_{ef\!f}$ and the dependence of
$\Delta E^A$ on $n_d$.}
\label{gef}
\end{figure}

Until now, we have presented results for arbitrary N. In the following, we
will focus on the limit $N\rightarrow \infty$ with $2S/N$ and $q/N$ finite.
In this limit the expressions obtained are simplified, and it is easier to
present the main features of the model. Thus, in the large-N limit, we have 
\begin{eqnarray}
J_{ef\!f} = -\left(\frac{4t^2}{JN}\right) \left[\left(\frac{N(N-2q)}{
q(N-q)}\right)\frac{1}{(2S+N-q)}\right]~.
\label{jef}
\end{eqnarray}
$J_{ef\!f}$ is of order $(1/N^2)$. The result expressed in Eq. (\ref{jef}) means that the
value of the effective coupling $J_{ef\!f}$ is independent of the number of 
conduction
electrons, $n_d$, on site 1 and coincides with
the result obtained in Ref. \onlinecite{cpt2} in the case $n_d=1$. 
This is due to the cancellation of the $(2S+n_d-1)$
factor that we have mentioned before.
The coupling remains ferromagnetic as long as
$q < N/2$, as can be seen in Fig. \ref{gef} and by inspection of the numerator in
the r.h.s. of Eq. (\ref{jef}).
As soon as $q>N/2$, the strong coupling fixed point becomes unstable.

\begin{figure}[ht]
\includegraphics[width=8.0cm]{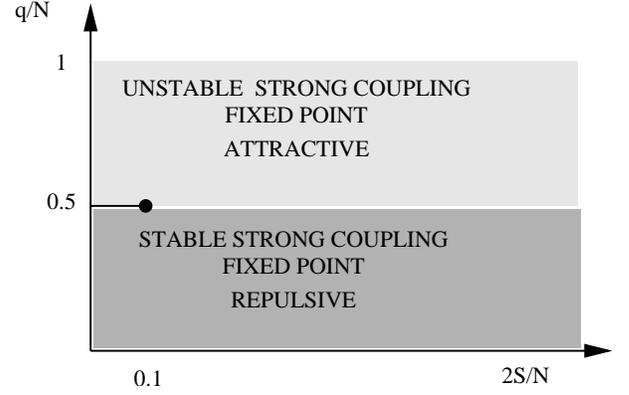}
\caption {Phase diagram of the model (for $2S/N$ finite), as a function of
the impurity parameters, $2S$, and $q$. As soon as $q>N/2$
the strong coupling fixed point becomes unstable. For $q=N/2$, the strong
coupling fixed point remains stable only for moderate values of $2S/N$ 
(short line ending in a point).}
\label{diagr}
\end{figure}

The case $q=N/2$ requires particular attention, since the leading contribution to 
$J_{ef\!f}$ vanishes. Taking into account the whole expression for the effective
coupling, we find that the strong coupling fixed point for an impurity with $q=N/2$ 
is stable as long as the bosonic parameter, $S$ is smaller than a critical value
\begin{eqnarray*}
S^*=\frac{1}{4}\left(N\sqrt{\frac{2N}{N-1}}-(N-2)\right)~.
\end{eqnarray*}
In the large-N limit we have
\begin{eqnarray*}
S^* = \left(\frac{\sqrt{2}-1}{4}\right)N+\frac{4+\sqrt{2}}{8}+
O(1/N) \sim \frac{N}{10}~.~
\end{eqnarray*}
We see that the strong coupling fixed point at $q=N/2$ becomes unstable already for 
moderate values of $S$ (Fig. \ref{diagr}). 

It is worth studying the energy shift for each state separately. 
It is easy to show that 
\begin{eqnarray}
\Delta E_0^A = -\left(\frac{2t^2}{J}\right)&& \!\!\!\!\!\!\!\!\!\! \left[ 
\left(\frac{n_d}{N-1}\right)
\left(\frac{q-1}{N-q-Q/N}\right) \right. \nonumber \\  &+& \!\!\!\!\!\left. 
\left(1-\frac{n_d}{N-1}\right)
\left(\frac{N-q}{q+Q/N}\right) \right]~,~~~
\label{dea}
\end{eqnarray}
\begin{eqnarray}
\Delta E_0^S = \Delta E_0^A +(2S+n_d-1)\frac{J_{ef\!f}}{2}~.
\label{des}
\end{eqnarray}
Since $J_{ef\!f}$ is ${\cal O}(1/N^2)$ in the large-N limit, the energy 
difference 
$(\Delta E_0^S - \Delta E_0^A)$ is ${\cal O}(1/N)$, and both energy levels 
$(\Delta E_0^S$ and $\Delta E_0^A)$ have the same leading term in 
${\cal O}(1)$. 
Notice that when we consider the large-N limit in which $Q/N$ is finite, the antisymmetric 
energy shift, $\Delta E^A_0$ is almost independent of $2S$. The dependence of 
$\Delta E_0^S$ on S is mainly contained in the $J_{ef\!f}$ term, which will eventually determine
the stability of the system.

\begin{figure}[h]
\includegraphics[width=8.0cm]{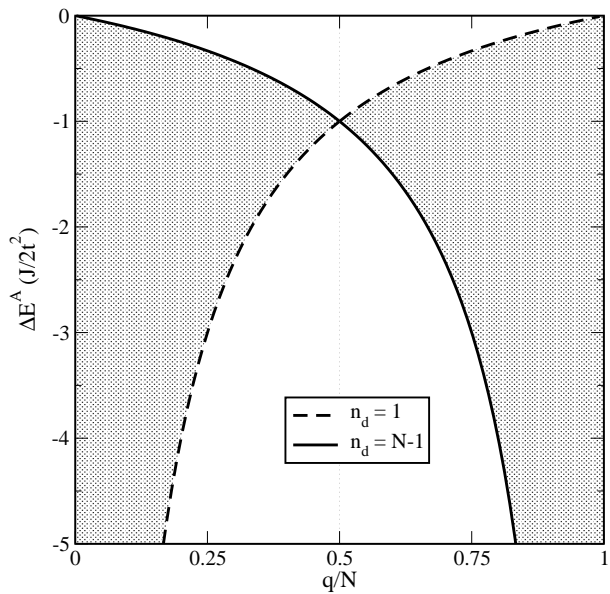}
\caption {Leading order term in the energy shift, $\Delta E_0^A \sim \Delta E_0^S$, as a function of $q/N$, 
for $1<n_d<(N-1)$ (shaded region), and in
the limiting cases  $n_d=1$ (dashed line), and  
$n_d=(N-1)$ (straight line). Notice that the
value at $q/N=1/2$ is equal to $-2t^2/J$, for any $n_d$. Note that for $q/N<0.5$, the energy is minimized for
$n_d=1$ while for $q/N>0.5$, the minimization is obtained for $n_d=(N-1)$}
\label{qpure}
\end{figure}

The most important property of Eq. (\ref{dea}) is its behavior as a function of $n_d$. To leading
order in $1/N$, we can write
\begin{eqnarray}
\Delta E_0^A = -\left(\frac{2t^2}{J}\right) \left[ 
\left(\frac{N-q}{q}\right) - \left(\frac{n_d}{N}\right)
\left(\frac{N(N-2q)}{q(N-q)}\right) 
\right]~.~
\label{dean}
\end{eqnarray}
Notice that the factor multiplying $n_d/N$ in the r.h.s. of Eq. (\ref{dean}) 
also appears in $J_{ef\!f}$ (see
Eq. (\ref{jef})). This result has the immediate following physical consequence.
The change of sign of $J_{ef\!f}$ -and hence of the stability of the strong coupling
fixed point- is directly connected to the change in the behavior of 
$\Delta E_0^A \sim \Delta E_0^S$ with $n_d$.
In particular, when $J_{ef\!f}=0$
\begin{eqnarray*}
\Delta E_0^A(q=N/2) = -\frac{2t^2}{J}~,
\end{eqnarray*}
for any value of $n_d$.
The value of $n_d$ that minimizes the energy given by Eq. (\ref{dean}) at 
arbitrary $q$ will depend on the sign of $(N-2q)$. In the regime where the
strong coupling fixed point is stable, 
$q/N < 1/2$, $J_{ef\!f}<0$, the lowest energy corresponds to $n_d= 1$, whereas 
for $q/N > 1/2$, $J_{ef\!f}>0$, the energy expressed in Eq. (\ref{dean}) is
minimized for $n_d=(N-1)$.
We have plotted $\Delta E_0^A$ in Fig. \ref{qpure} 
(compare to Fig. \ref{fdel}). The shaded region corresponds to 
the possible values of $\Delta E_0^A$ for the whole range of $n_d$, bounded 
by the limiting
cases, $n_d=1$, and $n_d=(N-1)$.
In the region where the strong coupling fixed point is stable, the electrons on site 1 are repelled by the
effective impurity (repulsive regime), whereas in the unstable, $q/N>1/2$ regime, the lowest energy in
second order perturbation theory corresponds to a state where the effective impurity
attracts as many electrons as possible on site 1 (attractive regime). This is precisely the mechanism behind
the two-stage quenching. The accumulation of electrons on site 1 is not related to $J_{ef\!f}$ which is
independent of $n_d$, but results from the dependence of $\Delta E_0^A \sim \Delta E_0^S$ with $n_d$.
Moreover, as we show in Appendix \ref{fermi}, $\Delta E_0^A$ coincides with the 
energy shift for a fermionic impurity (completely antisymmetric representation), in the large-N limit. 
When the impurity is fermionic, there is no degeneracy of the strong 
coupling fixed point, which is always stable, and the lowest order perturbation theory just shifts 
the ground state energy. Nevertheless, there are two regimes, repulsive and attractive, 
depending on the value of $q$, and characterized by the value of $n_d$ that minimizes 
the energy. This behavior is a consequence of the {\em particle-hole} symmetry in the
fermionic case, given by the transformations $q \rightarrow (N-q)$ and 
$n_d \rightarrow (N-n_d)$ (cf. Appendix \ref{fermi}).
The behavior of a fermionic impurity with $q$ is the same as in the case 
 $(N-q)$, if we reinterpret the electrons as holes and the impurity as made out of holes.
Therefore, if $n_d=1$ minimizes the energy for $q<N/2$ (electron repulsion), then the 
energy for a {\em hole} impurity, made out of $(N-q)$ fermions, is minimized by the state 
that repels the holes, $(N-n_d)=1$, implying an attraction of electrons, 
$n_d=(N-1)$.

The addition of a bosonic component to the impurity, leading to the formation of a row in 
the L-shaped Young tableau representing the impurity, breaks
this particle-hole symmetry. Whereas the two regimes described above are still present,
due to the fermionic component, the degeneracy of the states due to the bosonic component
leads to the instability of the strong coupling fixed point at the same point as where the 
the dressed impurity starts attracting the conduction electrons on site 1.

We finish by making some remarks on the physical properties of the model in the different
regimes. As is common to all models with an antiferromagnetic Kondo coupling, there will be 
a crossover from weak coupling above a given Kondo scale, $T_K$, to a low-energy regime.
When the strong coupling fixed point is stable, we should expect for $T\ll T_K$ a weak 
coupling of the effective impurity at
site-0 with the rest of the electrons. The physical properties at low temperature are
controlled by the degeneracy of the effective impurity, $d([2S-1])=C^{N-1}_{N+2S-2}$.
Thus, we should expect a residual entropy ${\cal S}^i \sim \ln C^{N-1}_{N+2S-2}$ and a
Curie susceptibility, $\chi^i \sim  C^{N-1}_{N+2S-2}/T$, with logarithmic corrections \cite{jaz,pgks}. 
This is the result that we would
expect for a purely symmetric impurity. The difference with respect to
the case at hand is that in
the L-shaped impurity model only $(N-q)$ electrons are allowed at the origin, instead of
$(N-1)$. Thus, we would expect to find different results for quantities that involve the
scattering phase shift of electrons off the effective impurity (see Appendix
\ref{fermi}).

In the $q>N/2$, we do not have access to the intermediate coupling fixed point that determines the low-energy
behavior. Nevertheless, it is reasonable to think that there would be a magnetic contribution
to the entropy, and a Curie-like contribution to the susceptibility, since the impurity
remains unscreened. This behavior is different from that of the multichannel
Kondo model, which is also characterized by an intermediate coupling fixed
point, but where the impurity magnetic degrees of freedom are completely
quenched \cite{aflw1}. The degeneracy of the true ground state is an open question, but we
can assume that the entropy will be smaller than that of the strongly coupled fixed point \cite{aflw2}.
It is in the scattering properties that we might be able to see the anomalous
features of this new fixed point more clearly.

\section{Conclusions}
In this paper, we have studied the $SU(N)$, single-channel Kondo model,
with a general impurity spin, involving both bosonic and fermionic 
degrees of freedom (corresponding, respectively, to the horizontal and
vertical directions in a L-shaped Young tableau). This model shows a
transition from a strong coupled fixed point to an intermediate coupling
fixed point, when
the amount of fermionic degrees of freedom, $q$, becomes larger than $N/2$.
We have identified the origin of this instability as due to 
the change, from repulsive to attractive, of the effective interaction 
between the dressed impurity and the conduction electrons in the 
neighboring sites. 
This change is already present in the
purely fermionic case, where it happens at the
particle-hole symmetry point, $q=N/2$. 

These results shed light on the
nature of the Kondo screening. 
Whereas the screening is restricted to 
a one-stage process in the strong-coupling regime, involving mainly the 
fermionic part of the impurity, the accumulation of conduction electrons 
on neighboring sites in the attractive regime gives rise to a two-stage
Kondo effect. The latter case corresponds to a regime where the low-energy physics is 
controlled by an intermediate-coupling fixed point. 
Obviously, the interesting open problem now is to understand the physics 
associated with the this new intermediate coupling fixed point. This
issue might have important future applications for the lattice problem,
with potential consequences for the understanding of non-Fermi liquid
behaviour observed in heavy-fermion systems.

\acknowledgements

The authors would like to thank N. Andrei for his continuous encouragement 
and discussions, and for a critical reading of this paper. We would also 
like to thank S. Burdin, P. Coleman, 
Ph. Nozi\`eres, et C. P\'epin for very helpful discussions. 
We are grateful to two different Programs, {\it Strongly Correlated Electron Systems} 
at the ICTP, Trieste in 1999, and at the Isaac Newton Institute for Mathematical Sciences,
Cambridge in 2000 where this work was initiated and further developed. We would 
like to take this opportunity to thank the organizers for creating a 
stimulating environment for scientific exchange.

\appendix 

\section{Composition of three fundamental representations of $SU(3)$}
\label{su3}

Before dealing with the general problem of constructing the highest weight
impurity states in $SU(N)$ in Appendix \ref{Lshape}, we will write down in
detail all the three-particle states with $SU(3)$ symmetry
\cite{lich}. This will allow
us to see how the states constructed with different numbers of bosons
and fermions can be basis
for representations with the same Young tableau. We will also see the role
of the $SU(3|3)$ and $SU(1|1)$ supersymmetry groups induced by the realization in
terms of bosons and fermions.

\begin{figure}[h]
\includegraphics[width=5.0cm]{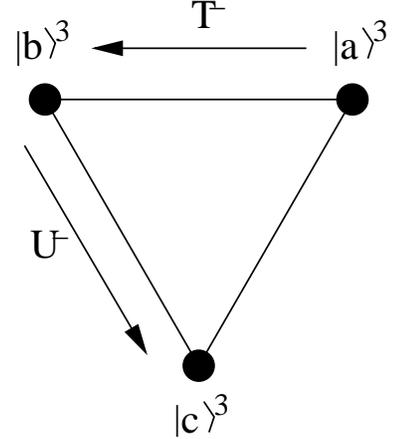}
\caption {Weight diagram for the fundamental representation of $SU(3)$, with the states and
the relevant lowering operators. }
\label{su30}
\end{figure}

\begin{figure*}[h]
\includegraphics[width=17.0cm]{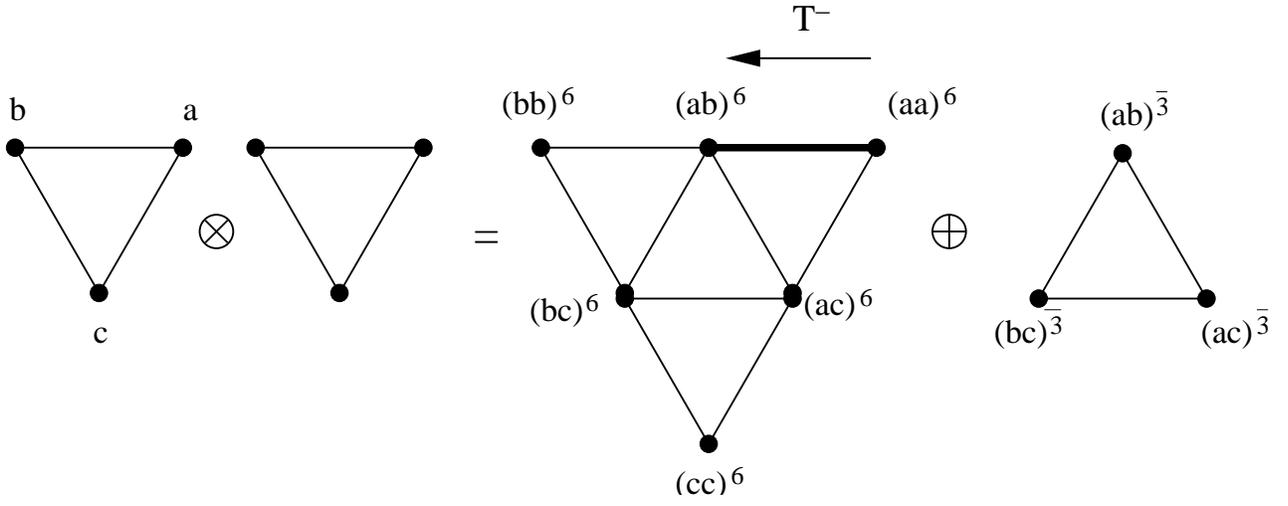}
\caption {Weight diagram for the Clebsch-Gordan series of the product of two fundamental 
representations of $SU(3)$.} 
\label{su31}
\end{figure*}

The direct product $\mathbf{3}\otimes \mathbf{3}\otimes \mathbf{3}$ of three
fundamental representations of $SU(3)$ gives the following Clebsh-Gordan
series 
\begin{equation}
\mathbf{3}\otimes \mathbf{3}\otimes \mathbf{3}=(\mathbf{3}\otimes \mathbf{6}%
)\oplus (\mathbf{3}\otimes \bar{\mathbf{3}})=\mathbf{10}\oplus \mathbf{8}%
^{1}\oplus \mathbf{8}^{2}\oplus \mathbf{1}~,~  \label{cgs}
\end{equation}%
where we identify each representation by its dimension. In terms of Young
tableaux, we have 
\begin{eqnarray*}
\yng(1)\otimes \yng(1)\otimes \yng(1)&=&\left( ~\yng(1)\otimes \yng(2)~\right)
\oplus \left( ~\yng(1)\otimes \yng(1,1)~\right)  \\ &=& \yng(3)~\oplus ~\yng
(2,1)~\oplus ~\yng(2,1)~\oplus ~\yng(1,1,1)~.
\end{eqnarray*}
The representations $\mathbf{6}$ and $\bar{\mathbf{3}}$ result from the
composition of two fundamental representations 
\begin{equation}
\mathbf{3}\otimes \mathbf{3}=\mathbf{6}\oplus \overline{\mathbf{3}}~.
\label{2par}
\end{equation}
In addition to Young Tableaux, we can use weight diagrams to describe the states in the
representation, Fig. \ref{su30}. In $SU(3)$, we associate a triangle to the fundamental 
representation, $\mathbf{3}$. Each vertex corresponds to a particular state of the multiplet, 
and the different states are related by the action of the lowering operators.
In Fig. \ref{su31} we include the weight diagrams associated with Eq. (\ref{2par}). 

The representation $\mathbf{6}$ is completely symmetric.
Its states are realized in terms of Schwinger
bosons. For instance its highest weight state can be written as 
\begin{equation*}
|aa\rangle _{b}^{\mathbf{6}}=\frac{1}{\sqrt{2!}}(b_{a}^{\dagger
})^{2}|0\rangle~,
\end{equation*}%
Here, and in the following, the values of $SU(3)$ spin are denoted by $a$, $b$,
and $c$.
Likewise, the representation $\overline{\mathbf{3}}$ is completely antisymmetric, and 
its states are more conveniently expressed in terms of fermions. For the highest weight
state we have 
\begin{equation*}
|ab\rangle _{f}^{\bar{\mathbf{3}}}=f_{a}^{\dagger }f_{b}^{\dagger }|0\rangle~.
\end{equation*}
There is another way of realize both $\mathbf{6}$ and $\bar{\mathbf{3}}$
using one boson and a fermion. Being symmetric, the highest weight of $\mathbf{6}$ is 
easy to write, since 
\begin{equation*}
|aa\rangle _{f}^{\mathbf{6}}=f_{a}^{\dagger }b_{a}^{\dagger }|0\rangle~,
\end{equation*}%
is already symmetrized. To obtain the highest weight of $\bar{\mathbf{3}}$,
we first have to find a state with the same quantum numbers in $\mathbf{6}$, by
acting with $T_{ab}^{-}=(f_{b}^{\dagger }f_{a}+b_{b}^{\dagger }b_{a})$ on 
$|aa\rangle _{f}^{\mathbf{6}}$, to get 
\begin{equation*}
|ab\rangle _{f}^{\mathbf{6}}=\frac{1}{\sqrt{2}}(f_{b}^{\dagger
}b_{a}^{\dagger }+f_{a}^{\dagger }b_{b}^{\dagger })|0\rangle~,
\end{equation*}%
and find a state orthogonal to $|ab\rangle _{f}^{\mathbf{6}}$ 
\begin{equation*}
|ab\rangle _{b}^{\bar{\mathbf{3}}}=\frac{1}{\sqrt{2}}(f_{b}^{\dagger
}b_{a}^{\dagger }-f_{a}^{\dagger }b_{b}^{\dagger })|0\rangle~.
\end{equation*}
This process is described in Fig. \ref{su31}.
It is easy to see that the states with the subindex $f$ are related to those
with the subindex $b$ by the $SU(1|1)$ supersymmetric operator $\theta
=\sum_{a}b_{a}^{\dagger }f_{a}$, so that $\theta |(\cdots )\rangle _{f}=%
\sqrt{2}~|(\cdots )\rangle _{b}$.
\begin{figure*}[h]
\includegraphics[width=17.0cm]{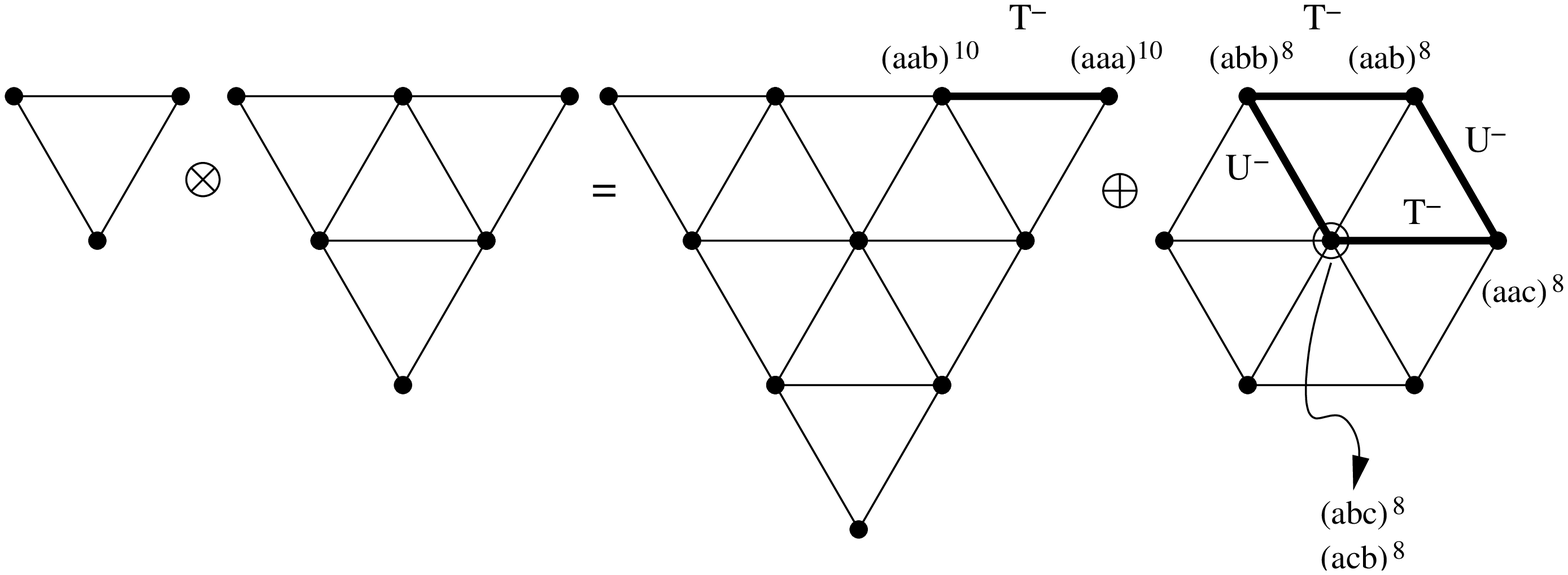}
\caption {Weight diagram for the Clebsch-Gordan series of the product $\mathbf{3} \otimes
\mathbf{6}$. We indicate some of the states and outline the process of obtaining
Clebsch-Gordan coefficients.} 
\label{su32}
\end{figure*}

Consider now the three-particle states. The easiest state to write is the
highest weight state in the most symmetric representation, $\mathbf{10}$
(cf. Fig. \ref{su32}) 
\begin{equation*}
|aaa\rangle ^{\mathbf{10}}=|a\rangle ^{\mathbf{3}}|aa\rangle ^{\mathbf{6}}~.
\end{equation*}%
It can be easily expressed in terms of bosons in agreement with Eq. (\ref{sbos})
\begin{equation*}
|aaa\rangle _{b}^{\mathbf{10}}=\frac{1}{\sqrt{3!}}(b_{a}^{\dagger
})^{3}|0\rangle~.
\end{equation*}%
Alternatively, we can use a realization with two bosons (from $\mathbf{6}$)
and a fermion (from $\mathbf{3}$) 
\begin{equation*}
|aaa\rangle _{f}^{\mathbf{10}}=\frac{1}{\sqrt{2!}}f_{a}^{\dagger
}(b_{a}^{\dagger })^{2}|0\rangle~.
\end{equation*}%
Other states of the representation are obtained by the repeated action of
lowering operators. For instance, 
\begin{equation}
|aab\rangle ^{\mathbf{10}}=\frac{1}{\sqrt{3}}(\sqrt{2}|a\rangle ^{\mathbf{3}%
}|ab\rangle ^{\mathbf{6}}+|b\rangle ^{\mathbf{3}}|aa\rangle ^{\mathbf{6}})~,
\label{10}
\end{equation}%
which leads to 
\begin{equation*}
|aab\rangle _{b}^{\mathbf{10}}=\frac{1}{\sqrt{2!}}(b_{a}^{\dagger
})^{2}b_{b}^{\dagger }|0\rangle~,
\end{equation*}
\begin{equation*}
|aab\rangle _{f}^{\mathbf{10}}=\frac{1}{%
\sqrt{6}}(2f_{a}^{\dagger }b_{a}^{\dagger }b_{b}^{\dagger }+f_{b}^{\dagger
}(b_{a}^{\dagger })^{2})~.
\end{equation*}

\begin{figure*}[h]
\includegraphics[width=17.0cm]{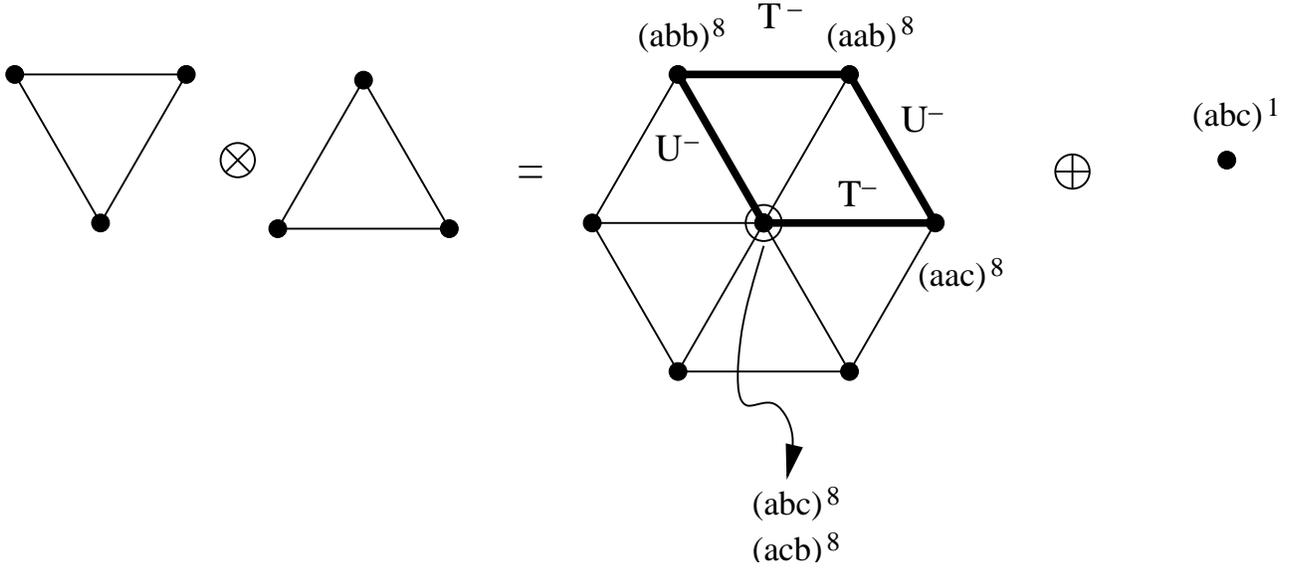}
\caption {Weight diagram for the Clebsch-Gordan series of the product $\mathbf{3} \otimes
\bar{\mathbf{6}}$.} 
\label{su33}
\end{figure*}

The octets, $\mathbf{8}$ are mixed symmetry representations that have to be
built by a combination of fermions and bosons. The Clebsch-Gordan series (\ref{cgs}) 
indicates that $\mathbf{8}^{1}$, built from the product $\mathbf{3}\otimes \mathbf{6}$ 
is naturally realized by states with one fermion and
two bosons, Fig. \ref{su32}, whereas $\mathbf{8}^{2}$ is realized by 
the product of one boson
and two fermions ($\mathbf{3}\otimes \bar{\mathbf{3}}$), Fig. \ref{su33}.
The highest weight state, $|aab\rangle ^{\mathbf{8}^{1}}$, of the octet 
$\mathbf{8}^{1}$ is orthogonal to $|aab\rangle ^{\mathbf{10}}$ defined in Eq. (\ref{10}) 
\begin{eqnarray*}
|aab\rangle _{b}^{\mathbf{8}^{1}}&=&\frac{1}{\sqrt{3}}(|a\rangle ^{\mathbf{3}}
|ab\rangle ^{\mathbf{6}}-\sqrt{2}|b\rangle ^{\mathbf{3}}|aa\rangle ^{%
\mathbf{6}})  \\ &=& \frac{1}{\sqrt{3}}b_{a}^{\dagger }(f_{a}^{\dagger
}b_{b}^{\dagger }-f_{a}^{\dagger }b_{b}^{\dagger })|0\rangle~,
\end{eqnarray*}%
in agreement with the general expression of the state $\psi _{b}$ given in 
(\ref{psib}). 

As usual, the other states of the octet are built by the
repeated action of generators $T^{-}=(f_{b}^{\dagger }f_{a}+b_{b}^{\dagger
}b_{a})$, $U^{-}=(f_{b}^{\dagger }f_{c}+b_{b}^{\dagger }b_{c})$, and $%
V^{-}=(f_{a}^{\dagger }f_{c}+b_{a}^{\dagger }b_{c})$, of $SU(3)$. For
instance 
\begin{eqnarray*}
\lefteqn{|abc\rangle _{b}^{\mathbf{8}^{1}} =
\frac{1}{\sqrt{2}}T^{-}|aac\rangle^{\mathbf{8}^{1}} } \\ 
&=& \frac{1}{\sqrt{6}}(|a\rangle ^{3}|bc\rangle ^{\mathbf{6}%
}+|b\rangle ^{\mathbf{3}}|ac\rangle ^{\mathbf{6}}-2|c\rangle ^{\mathbf{3}%
}|ab\rangle ^{\mathbf{6}}) \\
&=&\frac{1}{\sqrt{6}}(f_{a}^{\dagger }b_{b}^{\dagger }b_{c}^{\dagger
}+f_{b}^{\dagger }b_{a}^{\dagger }b_{c}^{\dagger }-2f_{c}^{\dagger
}b_{a}^{\dagger }b_{b}^{\dagger })|0\rangle~.
\end{eqnarray*}%
This state is degenerate, since there is another state in the multiplet with
the same quantum numbers. In order to find this last state, $|acb\rangle ^{%
\mathbf{8}^{1}}$, we have to combine the action of lowering operators with
orthogonality with respect to $|abc\rangle ^{\mathbf{8}^{1}}$. Acting with 
$U^{-}$ on $|abb\rangle ^{\mathbf{8}^{1}}$ leads to a state that is
not orthogonal to $|abc\rangle ^{\mathbf{8}^{1}}$. Therefore, we write 
\begin{equation*}
U^{-}|abb\rangle ^{\mathbf{8}^{1}}=\sqrt{2}(\alpha ~|abc\rangle ^{\mathbf{8}%
^{1}}+\beta ~|acb\rangle ^{\mathbf{8}^{1}})~.
\end{equation*}%
There are two ways of reaching states with quantum numbers $\{abc\}$
starting from $|aab\rangle ^{\mathbf{8}^{1}}$. We use this and the fact that 
$[T^{+},U^{-}]=0$, ($T^{+}=(T^{-})^{\dagger }$), to derive $\alpha $ 
\begin{eqnarray*}
2~\alpha &=&\sqrt{2}~{}^{\mathbf{8}^{1}}\langle abc|U^{-}|abb\rangle ^{%
\mathbf{8}^{1}}={}^{\mathbf{8}^{1}}\langle aac|T^{+}U^{-}T^{-}|aab\rangle ^{%
\mathbf{8}^{1}} \\
&=&{}^{\mathbf{8}^{1}}\langle aab|U^{+}T^{+}U^{-}T^{-}|aab\rangle ^{\mathbf{8%
}^{1}} \\ &=& 
{}^{\mathbf{8}^{1}}\langle aab|T^{+}T^{-}U^{+}U^{-}|aab\rangle ^{%
\mathbf{8}^{1}}=1~.
\end{eqnarray*}%
Hence one deduces $\alpha =1/2$ and $\beta =\sqrt{3}/2$ and one gets the
expression of the last state $|acb\rangle _{b}^{\mathbf{8}^{1}}$of the octet 
\begin{eqnarray*}
|acb\rangle _{b}^{\mathbf{8}^{1}}&=&\frac{1}{\sqrt{2}}(|a\rangle ^{\mathbf{3}%
}|bc\rangle ^{\mathbf{6}}-|b\rangle ^{\mathbf{3}}|ac\rangle ^{\mathbf{6}})
\\ &=&
\frac{1}{\sqrt{2}}(f_{a}^{\dagger }b_{b}^{\dagger }b_{c}^{\dagger
}-f_{b}^{\dagger }b_{a}^{\dagger }b_{c}^{\dagger })|0\rangle~.
\end{eqnarray*}

To summarize, the rule is to describe the highest weight state of the octet 
$|aab\rangle_b$ using the bosonic representation, $\psi _{b}$, and hence to
derive the other states of the octet by the above construction allowing to
recover the Clebsch-Gordan coefficients involved in the spin composition
related to the direct product $3\otimes 6$.

The highest weight state of the octet $\mathbf{8}^{2}$ is also the highest
weight state of $\mathbf{3}\otimes \bar{\mathbf{3}}$ 
\begin{equation*}
|aab\rangle _{f}^{\mathbf{8}^{2}}=|a\rangle ^{\mathbf{3}}|ab\rangle ^{\bar{%
\mathbf{3}}}=b_{a}^{\dagger }f_{a}^{\dagger }f_{b}^{\dagger }|0\rangle~,
\end{equation*}%
in agreement with the general expression of the state $\psi _{f}$, 
(\ref{psiff}). The construction of the other states follow the same lines as in
the case of the $\mathbf{8}^{1}$ octet. For instance, 
\begin{eqnarray*}
|abc\rangle _{f}^{\mathbf{8}^{2}} &=& 
\frac{1}{\sqrt{2}}(|a\rangle ^{\mathbf{3}%
}|bc\rangle ^{\bar{\mathbf{3}}}-|b\rangle ^{\mathbf{3}}|ca\rangle ^{\bar{%
\mathbf{3}}}) \\ &=& \frac{1}{\sqrt{2}}(b_{a}^{\dagger }f_{b}^{\dagger
}f_{c}^{\dagger }-b_{b}^{\dagger }f_{c}^{\dagger }f_{a}^{\dagger })|0\rangle~,
\end{eqnarray*}%
\begin{eqnarray*}
|acb\rangle _{f}^{\mathbf{8}^{2}}&=&\frac{1}{\sqrt{6}}(|a\rangle ^{\mathbf{3}%
}|bc\rangle ^{\bar{\mathbf{3}}}+|b\rangle ^{\mathbf{3}}|ca\rangle ^{\bar{%
\mathbf{3}}}-2|c\rangle ^{\mathbf{3}}|ab\rangle ^{\bar{\mathbf{3}}})
\\ &=& \frac{1%
}{\sqrt{6}}(b_{a}^{\dagger }f_{b}^{\dagger }f_{c}^{\dagger }+b_{b}^{\dagger
}f_{c}^{\dagger }f_{a}^{\dagger }-2b_{c}^{\dagger }f_{a}^{\dagger
}f_{b}^{\dagger })|0\rangle~.
\end{eqnarray*}
Once again, the two basis of states corresponding to the same Young tableau,
are related by the $SU(1|1)$ operators, $\theta^\dagger$, and $\theta$.

Finally, the singlet state, $|abc\rangle^{\mathbf{1}}$, is built by
orthogonality with the states $|abc\rangle^{\mathbf{8}^{2}}$, and $%
|acb\rangle^{\mathbf{8}^{2}}$, from the octet in the product $\mathbf{3}%
\otimes \bar{\mathbf{3}}$ (Fig. \ref{su33}) , 
\begin{equation}
|abc\rangle^{\mathbf{1}} =\frac{1}{\sqrt{3}}(|a\rangle^{\mathbf{3}}
|bc\rangle^{\bar{\mathbf{3}}}+|b\rangle^{\mathbf{3}}|ca\rangle^ {\bar{%
\mathbf{3}}}+|c\rangle^{\mathbf{3}}|ab\rangle^{\bar{\mathbf{3}}})~,
\label{singlet}
\end{equation}
in agreement with the expression of the states in the completely
antisymmetric representation of the spin (\ref{ferm}).
The simplest way to realize this state is with three fermions. Then 
\begin{equation}
|abc\rangle^{\mathbf{1}}_f= f_{a}^{\dagger }f_{b}^{\dagger}f_{c}^{\dagger
}\left\vert 0\right\rangle~.  \label{fff}
\end{equation}
But it can also be written with one boson and two fermions, either by acting
with $\theta$ on (\ref{fff}) or by substituting on (\ref{singlet}) 
\begin{equation*}
|abc\rangle^{\mathbf{1}}_b= \frac{1}{\sqrt{3}}(b^\dagger_af^\dagger_bf^%
\dagger_c+
b^\dagger_bf^\dagger_cf^\dagger_a+b^\dagger_cf^\dagger_af^\dagger_b)~.
\end{equation*}

We have shown how to construct the states for different representations of $SU(3)$.
These results can be summarized as tables of Clebsch-Gordan coefficients like the
ones that we presented in Tables \ref{t1}-\ref{bigt} for $SU(N)$.

\begin{figure}[h]
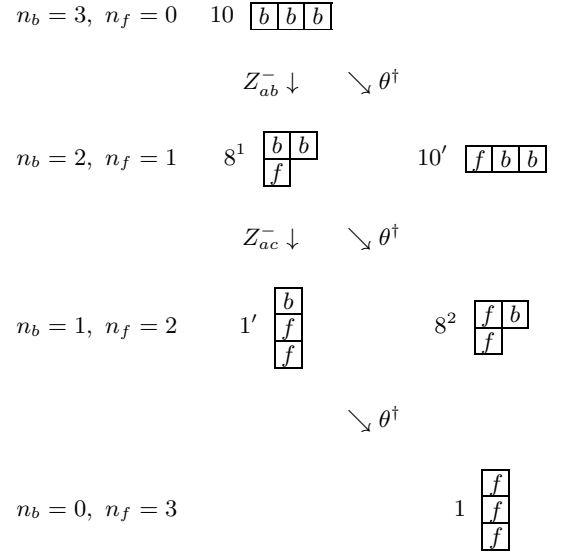

\[
\begin{array}{lccc}
n_{b}=3,~n_{f}=0~~ & 10~~\young(bbb) &  &  \\ 
&  &  &  \\ 
& Z_{ab}^{-}\downarrow & \searrow \theta ^{\dagger } &  \\ 
&  &  &  \\ 
n_{b}=2,~n_{f}=1~~ & 8^{1}~~\young(bb,f) &  & 10^{\prime }~~\young(fbb) \\ 
&  &  &  \\ 
& Z_{ac}^{-}\downarrow & \searrow \theta ^{\dagger } &  \\ 
&  &  &  \\ 
n_{b}=1,~n_{f}=2~~ & 1^{\prime }~~\young(b,f,f) &  & 8^{2}~~\young(fb,f) \\ 
&  &  &  \\ 
&  & \searrow \theta ^{\dagger } &  \\ 
&  &  &  \\ 
n_{b}=0,~n_{f}=3~~ &  &  & 1~~\young(f,f,f)%
\end{array}%
\]%
\caption {States with $Q=(n_{b}+n_{f})$ grouped according to $SU(3)$
representations}
\label{zth}
\end{figure}

Let us make some comments about the number of states. The direct product of
three fundamental representations of $SU(3)$ generates a space of dimension
27, which breaks down as a direct sum of irreducible representations
according to the Clebsch-Gordan series (\ref{cgs}). By considering all the
realizations of these states in terms of bosons and fermions subject only to
the constraint $Q=n_{f}+n_{b}=3$, we are working on the higher dimensional
space of a representation of $SU(3|3)$ with a
total of $38$ states, as represented in Fig \ref{zth} .The figure also
reports the relation between these states with the introduction of an
additional supersymmetric operator, $Z_{ar_{q}}^{-}=f_{r_{q}}^{\dagger }b_{a}$
, acting on the highest weight states. The states $\mathbf{10}$ and 
$\mathbf{10}^{\prime }$ respectively are identical (as well as $\mathbf{1}$ and 
$\mathbf{1}^{\prime }$) as far as the $SU(3)$ symmetry is concerned. This is 
not the case of the two states 
$\mathbf{8}^{1}$ and $\mathbf{8}^{2}$ which correspond to different spin
representations even if the associated Young tableaux are the same.
Altogether, one recovers the expected total of 27 different states. 

\section{Impurity states and low-lying states in the strong-coupling limit}
\label{apu}

We present in this appendix the explicit form of the highest weight (spin)
states for the impurity and for the low-lying states in the $J\rightarrow \infty$
limit. We discuss in detail the use of Young Tableaux both to describe the 
symmetry properties of the states and to study de $SU(N)$ generalization of the
composition of several spins.

\subsection{Impurity State}
\label{Lshape}

Before studying the general case of a L-shaped Young tableau representation
of spin, we will consider the two limiting cases of a completely symmetric
(bosonic) and antisymmetric (fermionic) representations of the spin in 
$SU(N)$.

The case of a completely symmetric representation of the spin is equivalent
to a system of $2S$ identical particles symmetric under the permutation of
two of them. The associated Young tableau is made of a single line of $2S$
boxes 
\begin{equation*}
\overbrace{\yng(5)}^{2S}\longleftrightarrow \sum_{P\in S_{2S}}P~,
\end{equation*}
expressed in shorthand notation as $[2S]$. Associated with the Young tableau,
there is a symmetrizer operator made out of the sum of all the permutations
of $2S$ elements. It is convenient to use an explicit representation of the
localized spin in terms of $N$ species of Schwinger bosons $b_{\alpha}$, 
$(\alpha=a,b,...,r_N)$ subject to the constraint 
\begin{equation*}
\hat{n}_{b}=\sum_{\alpha}b_{\alpha}^{\dagger }b_{\alpha}=2S~.
\end{equation*}
The $(N^{2}-1)$ components of the spin operator can be represented as 
$\mathbf{S}^{A}=\sum_{\alpha\beta}b_{\alpha}^{\dagger }
\boldsymbol{\tau}_{\alpha\beta}b_{\beta}$, while
the highest weight state (the analog of the state with the largest value of 
$S^{z}$ in $SU(2)$), can be written as 
\begin{equation}
|(a)^{2S}\rangle ^{\lbrack 2S]}=\frac{1}{\sqrt{(2S)!}}(b_{a}^{\dagger
})^{2S}|0\rangle~,
\label{sbos}
\end{equation}
where $|0\rangle $ denotes the vacuum state for the bosons. Other states of
the representation can be obtained from this one by the repeated action of
lowering operators, taking advantage of the underlying $SU(2)$ subalgebras
within $SU(N)$. Take, for instance, the $(2S+1)$ states $\{|(a)^{x}(b)^{y}
\rangle ^{\lbrack 2S]}\}$ with $x+y=2S$. They transform as a regular, $SU(2)$, 
spin-$S$ multiplet under the action of the $SU(N)$ operators 
$T_{ab}^{-}=b_{b}^{\dagger }b_{a}$, $T_{ab}^{+}=b_{a}^{\dagger }b_{b}$, and 
$T_{ab}^{z}=(b_{a}^{\dagger }b_{a}-b_{b}^{\dagger }b_{b})/2$. In particular, 
\begin{equation*}
T_{ab}^{-}~|a^{2S}\rangle ^{\lbrack 2S]}=\sqrt{2S}~|(a)^{2S-1}b\rangle
^{\lbrack 2S]}~,
\end{equation*}
or, in terms of bosons,
\begin{equation*}
|(a)^{2S-1}b\rangle ^{^{\lbrack 2S]}}=\frac{1}{\sqrt{(2S-1)!}}%
(b_{a}^{\dagger })^{(2S-1)}b_{b}^{\dagger }\left\vert 0\right\rangle.
\end{equation*}
Note that each index in the set of quantum numbers, $\{\alpha ,\beta ...,
\rho_{2S}\}$, describing the states of the representation, can take N values
independently of the rest of the set, and that for each set of values there
is only one state. The dimension of the representation is thus given by 
$C_{N+2S-1}^{2S}$, corresponding to the number of ways of choosing $2S$
elements out of a group of $(N+2S-1)$.

The other limiting case corresponds to a completely antisymmetric
representation of the spin. It is equivalent to the case of $q$ identical
particles antisymmetric under the permutation of two of them.\ The
associated Young tableau is made out of a single column of $q$ boxes 
\begin{equation*}
q\left\{ \yng(1,1,1,1,1)\right. ~\longleftrightarrow \sum_{P\in S_{q}}\delta
_{P}P~,
\end{equation*}
and expressed in shorthand notation as $[1^{q}]$ with $q<N$. Associated with
the Young tableau, there is an antisymmetrizer operator made out of the sum
of all the permutations of $q$ elements weighted by a $\delta _{P}=$ $\pm 1$
factor as for antisymmetric identical particles. It is convenient to use an
explicit representation of the localized spin in terms of $N$ species of
Abrikosov pseudo-fermions $f_{\alpha}$ $(\alpha=a,b,...,r_N)$ subject to 
the constraint 
\begin{equation*}
\hat{n}_{f}=\sum_{\alpha}f_{\alpha}^{\dagger }f_{\alpha}=q~.
\end{equation*}

The generators of $SU(N)$ in this realization are 
$\mathbf{S}^{A}=\sum_{\alpha\beta}f_{\alpha}^{\dagger }
\boldsymbol{\tau}_{\alpha\beta}^{A}f_{\beta}$, and the highest weight
state of the representation can be written as 
\begin{equation}
|ab...r_{q}\rangle ^{\lbrack 1^{q}]}=f_{a}^{\dagger }f_{b}^{\dagger
}...f_{r_{q}}^{\dagger }|0\rangle
\label{ferm}
\end{equation}
involving\ a set $\{a,b,...r_{q}\}$ of $q$ different indices. Other states
of the representation can be obtained from this one by the repeated action
of lowering operators such as $T_{ab}^{-}=f_{b}^{\dagger }f_{a}$, taking
advantage of the underlying $SU(2)$ subalgebras. Note that the states are $%
SU(2)$ doublets with respect to these subalgebras. The dimension of the
representation is given by $C_{N}^{q}$.

An important property of both kinds of representations is that the states
are non-degenerate. That is, each set of allowed quantum numbers completely
determine the state. This is not the case for mixed symmetry
representations, as we are about to see.
\smallskip

\begin{figure}[h]
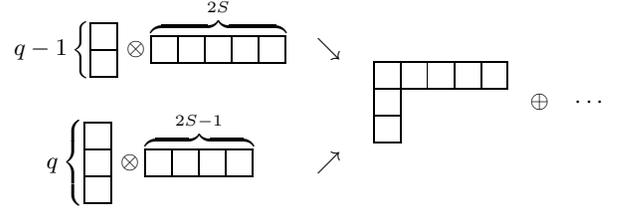

\label{tbp}
\begin{eqnarray*}
\begin{array}{cc}
q-1 \left\{\yng(1,1)\right.\otimes \overbrace{\yng(5)}^{2S} & ~\searrow \\ 
&  \\ 
q \left\{\yng(1,1,1)\right. \otimes \overbrace{\yng(4)}^{2S-1} & ~\nearrow%
\end{array}
~~~ \yng(5,1,1)~~\oplus~~ \cdots
\end{eqnarray*}
\caption {\label{yt2} Two ways of obtaining a L-shaped representation
 out of the direct product of a symmetric and an antisymmetric  $SU(N)$
representations}
\end{figure}

Let us now consider the general L-shaped representation of spin, 
Fig. \ref{yt1}, which interpolates between the previous two limits. 
Its dimension can be easily obtained using Robinson's formula \cite{ivs}:
the result is $\left(\frac{2S}{2S+q-1}\right)C_{N+2S-1}^{2S}C_{N-1}^{q-1}$
(Table \ref{t1}). 
The L-shaped representation is the result of the direct product of
a symmetric and an antisymmetric representation. This can be done in two,
non-equivalent ways (Fig. \ref{yt2}) : either as the $[1^{q-1}]\otimes
\lbrack 2S]\rightarrow \lbrack 2S,1^{q-1}]\oplus \cdots $ Clebsch-Gordan
series, or as $[1^{q}]\otimes \lbrack 2S-1]\rightarrow \lbrack
2S,1^{q-1}]\oplus \cdots $. The construction of the highest weight states
for each of the cases is detailed at the end of this subsection (cf. Eqs. 
(\ref{we1}) and (\ref{we2})) and leads to
\begin{eqnarray}
\psi _{b}&=&|(a)^{2S}b...r_{q}\rangle ^{\lbrack 2S,1^{q-1}]}
\nonumber \\
&=& \frac{1}{\sqrt{2S+q-1}}\frac{(b_{a}^{\dagger })^{2S-1}}
{\sqrt{(2S-1)!}}\mathcal{A}(b_{a}^{\dagger }
f_{b}^{\dagger }f_{c}^{\dagger }...f_{r_{q}}^{\dagger
})|0\rangle~,~~~~~~
\label{psib}
\end{eqnarray}
and
\begin{eqnarray}
\psi _{f}&=&|(a)^{2S}b...r_{q}\rangle ^{\lbrack 2S,1^{q-1}]}
\nonumber \\
&=& \frac{(b_{a}^{\dagger })^{2S-1}}{\sqrt{(2S-1)!}}
(f_{a}^{\dagger }f_{b}^{\dagger
}f_{c}^{\dagger }...f_{r_{q}}^{\dagger })|0\rangle~,~~~~
\label{psiff}
\end{eqnarray}
where $\mathcal{A}(\cdots )$ is the antisymmetrizer. 
The impurity spin operator has a
form which is independent of either $2S$ or $q$ and is 
given by $S=\sum_{\alpha\beta}(b_{\alpha}^{
\dagger }{\boldsymbol{\tau}}_{\alpha\beta}
b_{\beta}+f_{\alpha}^{\dagger }
{\boldsymbol{\tau}}_{\alpha\beta}f_{\beta})$. 
The first constraint has to do with the conservation of the number 
of particles \begin{equation}
\hat{Q}=(\hat{n}_{f}+\hat{n}_{b})=(2S+q-1)~,  \label{qc}
\end{equation}
where $(2S+q-1)$ denotes the number of boxes in the L-shaped Young tableau. 
In the limiting cases discussed
previously, once the value of $Q$ is fixed, the representation is completely
determined. Here, however, it is necessary to add a second constraint to
identify states with the right symmetry. To that end, it is worth noticing that the set
of states $\psi _{b}$ and$~\psi _{f}$ form a basis for a representation of
the larger, supersymmetric group $SU(N|N)$, with generators given as linear
combinations of the operators $b_{\alpha }^{\dagger }b_{\beta }$, $f_{\alpha
}^{\dagger }f_{\beta }$, $b_{\alpha }^{\dagger }f_{\beta }$, $f_{\alpha
}^{\dagger }b_{\beta }$. Thus, all the L-shaped impurities that interpolate
between the symmetric and the antisymmetric case are related by the
supersymmetric group. As a matter of fact, the constraints that fix the 
$SU(N)$ representation are obtained from operators in $SU(N|N)$ 
diagonal in spin, such as $\hat{n}_{f}$, $\hat{n}_{b}$, $\theta
=\sum_{\alpha }b_{\alpha }^{\dagger }f_{\alpha }$, and $\theta ^{\dagger }$.
Consider for instance, the action of $\theta ^{\dagger }$ on $\psi _{b}$, 
\begin{eqnarray*}
\lefteqn{
\theta ^{\dagger }~(b_{a}^{\dagger })^{2S-1}\mathcal{A}(b_{a}^{\dagger
}f_{b}^{\dagger }f_{c}^{\dagger }\dots f_{r_{q}}^{\dagger })~|0\rangle}
&& \\
&=& (2S+q-1)(b_{a}^{\dagger })^{2S-1}(f_{a}^{\dagger }f_{b}^{\dagger
}f_{c}^{\dagger }\cdots f_{r_{q}}^{\dagger })~|0\rangle~ .
\end{eqnarray*}
where the right hand side corresponds to $\psi _{f}$. This leads to the
relations 
\begin{eqnarray*}
\theta ^{\dagger }\psi _{b} &=&\sqrt{2S+q-1}~\psi _{f}~, \\
\theta ~\psi _{f} &=&\sqrt{2S+q-1}~\psi _{b}~.
\end{eqnarray*}%
Notice also that $\theta ^{\dagger }\psi _{f}=\theta \psi _{b}=0$. The
operators $\theta $ and $\theta ^{\dagger }$ relate states that transform
under a representation of $SU(N)$ given by the same Young Tableau. Together
with $\hat{Q}$, they form the $SU(1|1)$ supersymmetric algebra \cite{cpt1}
$\{\theta
,\theta ^{\dagger }\}=\hat{Q}$. 
Furthermore, the operators $P_{b}=\frac{1}{Q}%
\theta \theta ^{\dagger }$ and $P_{f}$ $=\frac{1}{Q}\theta ^{\dagger }\theta 
$ are the projectors out of the bosonic $\psi _{b}$ and the fermionic $\psi
_{f}$ states, respectively.
The states $\psi _{f}$ and $\psi _{b}$ are the exact analog of the familiar example
of the formation of the two octets, $\mathbf{8}^{1}$ and $\mathbf{8}^{2}$, 
out of the composition of three fundamental representations in $SU(3)$. In appendix \ref{su3}
we construct the states explicitely in this example and derive the
corresponding Clebsch-Gordan coefficients.

The second constraint
is then given by $\hat{\mathcal{Y}}$, a bilinear combination of the operators 
$\left\{ \hat{n}_{f},\hat{n}_{b},\theta ,\theta ^{\dagger }\right\}$ since it is 
a consequence of the invariance of the Casimir operator, $\hat{\mathcal{C}_{2}}$, 
(the $SU(N)$ generalization of $\boldsymbol{S}^2=S(S+1)$), which for a L-shaped 
representation is given by 
\begin{equation}
\mathcal{C}_{2}(\hat {R})=\sum_{A}\mathbf{S}^{A}\mathbf{S}^{A}=
\frac{1}{2}\left[ 
\hat{Q}(N-\frac{\hat{Q}}{N})-\hat{\mathcal{Y}}\right]~.
\label{c2r}
\end{equation}
Here, $\hat{\mathcal{Y}}=\hat{Q}(\hat{n}_{f}-\hat{n}_{b})+[\theta ,\theta
^{\dagger }]$. Once the first constraint, (\ref{qc}) is fulfilled, the
invariance of the Casimir $\hat{\mathcal{C}_{2}}$ is ensured provided that
the operator $\hat{\mathcal{Y}}$ is invariant too. This leads to the
second constraint 
\begin{equation}
\hat{\mathcal{Y}}=Q(q-2S)
\label{qys}
\end{equation}
It is easy to check that the operators $\theta $ and $\theta
^{\dagger }$ commute with $\hat{Q}$ and $\hat{\mathcal{Y}}$, which implies
that the constraints are also compatible with the $SU(1|1)$ supersymmetry.
Note that 
this was not the case for the operator $\hat{\mathcal{Y}}^{\prime }=\hat{n}_{f}
-\hat{n}_{b}+\frac{1}{Q}[\theta ,\theta ^{\dagger }]$ defined in Eq. 10 of 
ref. \onlinecite{cpt1}.

The constraints completely
determine the representation, but they cannot distinguish between the states 
$\psi _{f}$ and $\psi _{b}$. The physical properties of the system 
depend only on the
Young tableau associated with the Kondo impurity, and not on the particular
way the representation basis is constructed.
\bigskip

We finish this section by describing in detail the construction of the
relevant states of the impurity multiplet.

The direct product of irreducible representations of $SU(N)$ decomposes into
a direct sum of irreducible representations (Clebsch-Gordan series). A
well-known example is the addition of angular momentum in $SU(2)$. In order
to find the states in the new basis (leading to the Clebsch-Gordan
coefficients (CGC)), we follow a similar procedure to the one used for
angular momentum (see also Appendix \ref{su3}). That is, we first identify the highest 
weight state of the product of representations, which is always non-degenerate, with 
the highest
weight state of the most \emph{symmetric} of the representations in the
Clebsch-Gordan series. Then, we use lowering operators (particular
combinations of the generators of $SU(N)$) to generate other states in the
same representation. For arbitrary $N$ there might be more than one state
with the same quantum numbers in the same representation (as in the $SU(3)$ octet,
$\mathbf{8^1}$ and $\mathbf{8^2}$), and we will have to find orthogonal
combinations. 
Finally, we find states in the next
representation by looking for additional orthogonal states with
the same quantum numbers, and acting on them
with lowering operators.

The direct product of a symmetric and an antisymmetric representation of 
$SU(N)$ can be expressed as a direct sum of two L-shaped representations,
according to the Clebsh-Gordan series 
\begin{widetext}
\begin{eqnarray*}
\begin{array}{ccccccc}
[2S-1] & \otimes & [1^{q}] & = & [2S,1^{q-1}] & \oplus & [2S-1,1^{q}] \\ 
&  &  &  &  &  &  \\ 
\overbrace{\yng(5)}^{[2S-1]} & \otimes & q\left\{ \yng(1,1,1) \right. & = & 
q \left\{ \yng(6,1,1) \right. & \oplus & (q+1) \left\{ \yng(5,1,1,1) \right.%
\end{array}
\end{eqnarray*}
with dimensions 
\begin{eqnarray*}
C_{N+2S-2}^{2S-1}~ C_{N}^{q}= \frac{2S}{2S+q-1}~ C_{N+2S-1}^{2S}~
C_{N-1}^{q-1} + \frac{2S-1}{2S+q-1}~ C_{N+2S-2}^{2S-1}~ C_{N-1}^{q}~.
\end{eqnarray*}
\end{widetext}

We proceed now to write the states in each L-shaped representation. The 
\emph{most symmetric} (highest weight state) of the product $[2S-1]\otimes
[1^{q}]$ is 
\begin{eqnarray*}
|(a)^{2S-1}\rangle^{[2S-1]}|abc\dots r_{q}\rangle^{[1^{q}]}~,
\end{eqnarray*}
where $2S$ particles have the same quantum number $a$. This state is also
the highest weight state of the most symmetric of the representations, 
$[2S,1^{q-1}]$, 
\begin{eqnarray}
|(a)^{2S}bc\dots r_{q}\rangle^{[2S,1^{q-1}]} =
|(a)^{2S-1}\rangle^{[2S-1]}|abc\dots r_{q}\rangle^{[1^{q}]}~. &&
\nonumber \\ && \label{ta}
\end{eqnarray}
Notice that the 
lowering operators that transform the value $a$ into $\delta \in
[b,r_q]$ affect the $[2S-1]$ state only, since the $[1^{q}]$ term is
completely antisymmetric. Thus, these states are also non-degenerate 
\begin{eqnarray}
&& |(a)^{2S-1}bc\dots(\delta)^2\dots r_{q}\rangle^{[2S,1^{q-1}]} \nonumber 
\\ &&  = 
~|(a)^{2S-2}\delta\rangle^{[2S-1]}|abc\dots r_{q}\rangle^{[1^{q}]}~.
\label{we1}
\end{eqnarray}
Other states with $2S$ of the $\alpha_j$ equal to $a$, can be obtained from 
Eq. (\ref{ta}). They are also nondegenerate. For instance the state with the
value $\gamma$ replaced by $r_{q+1}$ is 
\begin{eqnarray}
&& |(a)^{2S}bcd\dots r_{q}r_{q+1}(no~\gamma)\rangle^{[2S,1^{q-1}]} \nonumber 
\\ && =
|(a)^{2S-1}\rangle^{[2S-1]}|abcd\dots r_{q}r_{q+1}(no~\gamma)
\rangle^{[1^{q}]} ~.~~  \label{ac}
\end{eqnarray}
Altogether, there are $q$ states of this type (if we restrict the $\alpha_j$
to the range $[a,r_{q+1}]$).

Next, we construct the states in $[2S,1^{q-1}]$ where $(2S-1)$ of the $%
\alpha_j $ are equal to $a$ and the rest of labels are different
and take values
in the range $[b,r_{q+1}]$. There are $q$ linearly independent states of
this kind. For instance, acting with $T_{ac}^-$ on (\ref{ac}) we get 
\begin{eqnarray}
\lefteqn{T_{ac}^- |(a)^{2S}bcd\dots
r_{q}r_{q+1}(no~\gamma)\rangle^{[2S,1^{q-1}]}} && \nonumber \\  &=& \sqrt{2S}~|(a)^{2S-1}bcd
\dots r_{q}r_{q+1}\rangle^{[2S,1^{q-1}]}   \nonumber \\
&=&  \sqrt{2S-1}~ |(a)^{2S-2}\gamma\rangle^{[2S-1]}|abc\dots r_{q}r_{q+1}(no~\gamma)
\rangle^{[1^{q}]}    \nonumber \\
&+& |(a)^{2S-1}\rangle^{[2S-1]}|bcd\dots r_{q}r_{q+1}
\rangle^{[1^{q}]}~.  \label{ga}
\end{eqnarray}
The orthogonalization of these $q$ states leads to the first $q$ rows of
Table \ref{bigt} of CGC. The last line in the table correspond to a state
with the same
quantum numbers ($(a)^{2S-1}bcd\dots r_qr_{q+1}$), and orthogonal to all the
states in $[2S,1^{q}]$. This state is the highest weight of $[2S-1,1^{q}]$.
It is easy to see that it must be of the form 
\begin{eqnarray}  
\lefteqn{|(a)^{2S-1}bcd\dots r_{q}r_{q+1}\rangle^{[2S-1,1^{q}]}} &&  \nonumber 
\\ &=&\frac{1}{%
\scriptstyle \sqrt{2S+q-1}}( \sqrt{2S-1}~|(a)^{2S-1}\rangle^{[2S-1]}|bcd%
\dots r_q\rangle^{[1^{q}]}  \nonumber \\
&+& \sum_{\beta=b}^{r_q}(-1)^{\delta_{a\beta}} |(a)^{2S-2}
\beta\rangle^{[2S-1]}|bc\dots r_q (no~\beta)\rangle^{[1^{q}]})~. \nonumber \\
&& \label{we2}
\end{eqnarray}

\subsection{Ground state}

The explicit form of the highest weight state for the ground state 
corresponds to Eq. (\ref{grst}),
\begin{equation*}
|GS\rangle _{\{a\}aa}^{[2S-1]}=\frac{1}{\scriptstyle\sqrt{(2S-1)!}}%
(b_{a}^{\dagger })^{2S-1}|\Delta \rangle~ , 
\end{equation*}%
with 
\begin{equation*}
|\Delta \rangle \equiv \frac{1}{\gamma }\mathcal{A}(b_{i_{1}}^{\dagger
}(\prod_{\alpha =i_{2}}^{i_{q}}f_{\alpha }^{\dagger })(\prod_{\beta
=i_{q+1}}^{i_{N}}c_{\beta }^{\dagger }))|0\rangle~ ,
\end{equation*}
\begin{equation*}
\gamma \equiv \sqrt{%
(2S+N-1)C_{N-1}^{q-1}}~.
\end{equation*}
Notice the additional term $C^{q-1}_{N-1}$ in the normalization factor 
$\gamma$, as compared to Eq. (\ref{psib}), due to the
presence of two kinds of fermions $f_{\alpha }^{\dagger }$ and $c_{\alpha
}^{\dagger }$. 
We adopt the bosonic realization of the impurity $\psi_b$, which simplifies
the calculations. We would like to emphasize that all the results are 
independent of the realization chosen, since only the form of the Young
tableau is relevant to the interaction.

Other 
states in the same
$[2S-1]$ multiplet can be obtained by just acting on the $(b_{a}^{\dagger })^{2S-1}$ 
term. For instance 
\begin{eqnarray}
\lefteqn{T^{-}~|GS\rangle _{\{a\}aa} = 
\sqrt{2S-1}~|GS\rangle _{\{a\}ab}} && \nonumber \\ &=& (2S-1)\frac{%
1}{\scriptstyle\sqrt{(2S-1)!}}~b_{b}^{\dagger }(b_{a}^{\dagger
})^{2S-2}|\Delta \rangle ~,
\label{22}
\end{eqnarray}
\begin{eqnarray}
U^{-}~|GS\rangle _{\{a\}ab} &=&|GS\rangle _{\{a\}ac} \nonumber \\
&=& \frac{1}{\scriptstyle%
\sqrt{(2S-2)!}}~b_{c}^{\dagger }(b_{a}^{\dagger
})^{2S-2}|\Delta \rangle ~,
\label{223}
\end{eqnarray}
and 
\begin{eqnarray}
|GS\rangle _{\{a\}ab}&=&\frac{1}{\scriptstyle\sqrt{(2S-2)!}}b_{b}^{\dagger
}(b_{a}^{\dagger })^{2S-2}|\Delta \rangle ~, \nonumber \\ 
|GS\rangle _{\{a\}bb} &=& \frac{%
1}{\scriptstyle\sqrt{2(2S-3)!}}(b_{b}^{\dagger })^{2}(b_{a}^{\dagger
})^{2S-3}|\Delta \rangle ~,  \nonumber \\
|GS\rangle _{\{a\}ac} &=& \frac{1}{\scriptstyle\sqrt{(2S-2)!}}b_{c}^{\dagger
}(b_{a}^{\dagger })^{2S-2}|\Delta \rangle ~, \nonumber \\
|GS\rangle _{\{a\}bc} &=& \frac{%
1}{\scriptstyle\sqrt{(2S-3)!}}b_{b}^{\dagger }b_{c}^{\dagger
}(b_{a}^{\dagger })^{2S-3}|\Delta \rangle ~.
\label{23}
\end{eqnarray}

\subsection{Excited states}
Let us now write the expression of the excited states of the strong-coupling
fixed point. 
The states in $|GS+1\rangle^S$ transform as the completely symmetric
representation $[2S]$. The highest weight state can be obtained by acting
with $c^\dagger_a$ on the ground state,
\begin{equation*}
|GS+1\rangle _{\{a\}aaa}^{S}=\frac{1}{\Omega}c_{a}^{\dagger
}|GS\rangle _{\{a\}aa}~,
\end{equation*}
where the normalization factor, $\Omega=\sqrt{\frac{2S+q-1}{2S+N-1}}$,
appears \cite{cpt2} because the additional c-electron has to be 
antisymmetrized with
respect to the $(N-q)$ electrons already present on site-0.
Other states in the multiplet can
be obtained by repeated action of the lowering operators, as in 
Eqs. (\ref{22},\ref{23}). For instance
\begin{eqnarray}
\lefteqn{|GS+1\rangle _{\{a\}aab}^{S} =} && \nonumber  \\ &&
\frac{1}{\Omega}\frac{1}{\sqrt{2S}}%
\left[ \sqrt{2S-1}c_{a}^{\dagger }|GS\rangle _{\{a\}ab}~+c_{b}^{\dagger
}|GS\rangle _{\{a\}aa}~\right] ~,~  \label{gsaab} 
\end{eqnarray}
\begin{eqnarray*}
|GS+1\rangle _{\{a\}abc}^{S} &=& \frac{1}{\Omega}\frac{1}{\sqrt{2S}}
\left[ \sqrt{2S-2}c_{a}^{\dagger }|GS\rangle _{\{a\}bc}  \right. \\
&+& \left. c_{b}^{\dagger
}|GS\rangle _{\{a\}ac}
~+ c_{c}^{\dagger }|GS\rangle _{\{a\}ab}\right]~.
\end{eqnarray*}
States in $|GS+1\rangle^A$ transform as $[2S-1,1]$. The highest weight
state is $|GS+1\rangle^A_{\{a\}aab}$, and it will be orthogonal to 
the state defined in Eq.(\ref{gsaab}). Thus
\begin{eqnarray*}
\lefteqn{|GS+1\rangle _{\{a\}aab}^{A}=} && \\ && \frac{1}{\Lambda}\frac{1}{\sqrt{2S}}
\left[ c_{a}^{\dagger }|GS\rangle _{\{a\}ab}~-\sqrt{2S-1}c_{b}^{\dagger
}|GS\rangle _{\{a\}aa}~\right]~,
\end{eqnarray*}
with \cite{cpt2} $\Lambda = \sqrt{\frac{q-1}{N-1}}$. Notice the difference with 
respect to the normalization factor. Only $\Omega$ depends on $2S$.

Other states in $|GS+1\rangle ^{A}$ are obtained in a very
similar way to the construction of the octets, $\mathbf{8^1}$
and  $\mathbf{8^2}$ in $SU(3)$ 
(see Appendix \ref{su3}). For
instance
\begin{eqnarray*}
|GS+1\rangle _{\{a\}abc}^{A} &=& \frac{1}{\Lambda}\frac{1}{\sqrt{%
2S(2S-1)}}\left[ \sqrt{2S-2}c_{a}^{\dagger }|GS\rangle _{bc}
\right. \\
 + c_{b}^{\dagger}
|GS\rangle _{ac} \!\! &-& \left. \!\! 
(2S-1)c_{c}^{\dagger }|GS\rangle _{ab}\right] ~,  
\end{eqnarray*}
\begin{eqnarray*}
\lefteqn{|GS+1\rangle _{\{a\}acb}^{A} =} && \\ 
&& \frac{1}{\Lambda}\frac{1}{\sqrt{2S-1%
}}\left[ c_{a}^{\dagger }|GS\rangle _{bc}~-\sqrt{2S-2}c_{b}^{\dagger
}|GS\rangle _{ac}\right]~.
\end{eqnarray*}

We collect the coefficients for the states with quantum numbers 
$\{\{a\}abc\}$ in Table \ref{su3cg} of Clebsch-Gordan coefficients 
corresponding to the direct product $[1]\otimes[2S-1]$ in $SU(N)$, 
following the notation of ref. \onlinecite{chwlw}.
\begingroup
\squeezetable
\begin{table}[h]
\caption {Clebsch-Gordan coefficients for the excited states with one
additional electron, corresponding to the product $[1]\otimes[2S-1]$, and
to states with quantum numbers $\{\{a\}abc\}$. The normalization factor
is $1/\sqrt{N}$, and a $\sqrt{~}$ is understood over each coefficient
\cite{chwlw}.
The minus sign indicates a negative sign in front of the square root,
$-\sqrt{~}$.}
\label{su3cg}
\begin{ruledtabular}
\begin{tabular}{c|c|ccc}  &&&& \\
& $N$ & $c^\dagger_a|GS\rangle_{\{a\}bc}$ & 
$c^\dagger_b|GS\rangle_{\{a\}ac}$ &
$c^\dagger_c|GS\rangle_{\{a\}ab}$
\\ &&&& \\  \hline &&&& \\ 
$\Omega|GS+1\rangle_{\{a\}abc}^S$ & $2S$ & $2S-2$ & 1 & 1
\\ &&&& \\
$\Lambda|GS+1\rangle_{\{a\}abc}^A$ & $2S(2S-1)$ & $2S-2$ & 1 & $-(2S-1)^2$
\\ &&&& \\
$\Lambda|GS+1\rangle_{\{a\}acb}^A$ & $2S-1$ & $1$ & $-(2S-2)$ & 0
\\ &&&& \\
\end{tabular} 
\end{ruledtabular}
\end{table}
\endgroup

We finish this section with the excited states corresponding to the 
multiplet $|GS-1\rangle$, characterized by one less conduction electron 
than in the ground state. They transform as $[2S-1,1^{N-1}]$, and the highest 
weight state is
\begin{equation*}
|GS-1\rangle=\frac{1}{\scriptstyle\sqrt{(2S-1)!}}(b_{a}^{\dagger
})^{2S-1}|\Delta ^{\prime }\rangle , 
\end{equation*}
with
 \begin{equation*}
|\Delta ^{\prime }\rangle \equiv \frac{1}{\gamma ^{\prime }}\mathcal{A}%
(b_{i_{1}}^{\dagger }(\prod_{\alpha =i_{2}}^{i_{q}}f_{\alpha }^{\dagger
})(\prod_{\beta =i_{q+1}}^{i_{N-1}}c_{\beta }^{\dagger }))|0\rangle~,
\end{equation*}
\begin{equation*}
\gamma ^{\prime }\equiv \sqrt{(2S+N-2)C_{N-1}^{q-1}}~. 
\end{equation*}
Here, $|\Delta'\rangle$ transforms as the $N$-dimensional, $[1^{N-1}]$,
representation of $SU(N)$.

\section{The case of the antisymmetric impurity}
\label{fermi}

The ground state for the strong coupling fixed point of a Kondo model with a fermionic
impurity (completely antisymmetric representation), $|GS\rangle^{[1^N]}$ is a singlet 
formed with $(N-q)$ conduction electrons at site 0, and has the same form 
as $|\Delta\rangle$, 
in (\ref{delta}), but with a different normalization factor
\begin{equation}
|GS \rangle^{[1^N]} \equiv \frac{1}{\sqrt{C_N^q}}\mathcal{A}(
(\prod_{\alpha =i_{1}}^{i_{q}}f_{\alpha }^{\dagger })(\prod_{\beta
=i_{q+1}}^{i_{N}}c_{\beta }^{\dagger }))|0\rangle~,
\label{gsq}
\end{equation}
The hopping hamiltonian leads to two types of processes, as described earlier, where the 
intermediate states have either one more or one less conduction 
electron. With the help of 
Table \ref{casi} we can write the excitation energies
\begin{eqnarray*}
\Delta E_1 = \frac{J}{2}\left(\frac{N+1}{N}\right)(N-q)~,~~~
\Delta E_2=\frac{J}{2}\left(\frac{N+1}{N}\right)q~. 
\end{eqnarray*}
It is clear that $\Delta E_1$ and $\Delta E_2$ are related by the {\em particle-hole} 
transformation $q \leftrightarrow (N-q)$. 

The ground state with $n_d$ electrons on site 1 is
\begin{eqnarray*}
|GS,n_d\rangle = (d^\dagger_a d^\dagger_b \cdots d^\dagger_u)|GS\rangle~.
\end{eqnarray*}  
Since there is only one intermediate state for each process, we can use the trick 
described in section \ref{trick}, and write
\begin{eqnarray}
M_1 = t^2 \left(n_d-\sum_{\sigma,\sigma'}\langle GS,n_d| c^\dagger_\sigma d^\dagger_{\sigma'}
d_{\sigma}c_{\sigma'}|GS,n_d\rangle\right)~,~
\label{m1}
\end{eqnarray} 
As long as $n_d>1$, the second term in Eq. (\ref{m1}) vanishes for $\sigma \ne \sigma'$. $c_\sigma$
acting on $|GS,n_d\rangle$ just counts the number of terms where there is a $c^\dagger_\sigma$.
There are $C^{q}_{N-1}$ such terms. Thus,
\begin{eqnarray*}
\sum_{\sigma,\sigma'}\langle GS,n_d| c^\dagger_\sigma d^\dagger_{\sigma'}
d_{\sigma}c_{\sigma'}|GS,n_d\rangle &=&  
n_d \left(\frac{C^q_{N-1}}{C^q_N}\right) \\ & =&  n_d \left(
\frac{N-q}{N}\right)~,
\end{eqnarray*}
and
\begin{eqnarray*}
M_1 = t^2 n_d \frac{q}{N}~.
\end{eqnarray*}
Finally, following Eq.(\ref{53}), 
\begin{eqnarray*}
M_2 = t^2(n_c-n_d)+M_1 = t^2(N-n_d)\left(\frac{N-q}{N}\right)~.
\end{eqnarray*}
$M_1$ and $M_2$ are related by the same transformations
$(q \leftrightarrow N-q,~n_d\leftrightarrow N-n_d)$ as the excitation energies. 
That means that the energy shift is
invariant under these transformations
\begin{eqnarray*}
\lefteqn{\Delta E_f = -\left(\frac{2t^2}{J}\right)} && \\ && \times 
\left(\frac{n_d}{N+1} \left(\frac{q}{N-q}\right)+
\frac{N-n_d}{N+1}\left(\frac{N-q}{q}\right)\right)~.
\end{eqnarray*}
In the large-N limit, this result is equivalent to Eq. (\ref{dean}). To leading order in $1/N$,
the energy shift of the strong coupling fixed point is determined by the fermionic component,
and the behavior under the particle-hole transformation.

We end this appendix by studying the phase shift $\delta$, of the conduction 
electrons scattered off the impurity site. This quantity characterizes 
the impurity contribution to the resistivity, $\rho^i$. At zero temperature and 
magnetic field, we have \cite{hew}
\begin{eqnarray}
\rho^i \propto \sin^2\delta~.
\end{eqnarray}
The phase shift for antisymmetric impurities in $SU(N)$, was computed in
ref. \onlinecite{pgks}. In the completely screened case it reads
\begin{eqnarray}
e^{2i\delta} = -e^{-i\pi\left(1-2\frac{q}{N}\right)}~.
\end{eqnarray}
If we choose the phase shift so that $|\delta|<\frac{\pi}{2}$, we have
\begin{eqnarray}
\delta=\left\{ \begin{array}{rl} \pi\left(\frac{q}{N}\right), & q<N/2 \\
					& \\
	                         -\pi\left(\frac{N-q}{N}\right), & q>N/2
		\end{array}
	\right.
\label{phase}
\end{eqnarray}
The unitary limit, $|\delta|=\frac{\pi}{2}$ is reached in the particle-hole
symmetric case, $q=N/2$. We see that this corresponds to the point where
$\Delta E_f$ is independent of $n_d$, indicating the change from the
attractive to the repulsive regime.

\section{Details of the calculations of the matrix elements}
\label{matrix}

Here we construct explicitely the excited states that are involved in
the second order perturbation theory, and then we 
compare them to the action of the hopping term,
 $(c_{\sigma }^{\dagger }d_{\sigma })$,
on the ground state. First, we add a $c$-electron to the ground state, and
then we combine it with $(n_{d}-1)$ electrons from site 1.

\subsection{Symmetric process, $|GS,n_d\rangle^{S}$}

The strong coupling excited state $|GS+1\rangle^{S}$ is easy to compute,
since it is the highest weight state in the product $[2S-1]\otimes[1]
\rightarrow [2S] \oplus [2S-1,1]$. We have 
\begin{eqnarray*}
|GS+1\rangle^{S}_{aaa} = \frac{1}{\Omega}c^\dagger_{a}|GS\rangle_{aa}~,
\end{eqnarray*}
with the normalization factor $\Omega^2=\frac{2S+q-1}{2S+N-1}$.
Other states within the
same representation, which transform as $[2S]$, can be obtained by acting with the 
corresponding lowering operators. For instance, 
\begin{eqnarray*}
|GS+1\rangle^{S}_{aab} = \frac{1}{\Omega\sqrt{2S}}\left( \sqrt{2S-1}%
~c^\dagger|GS\rangle_{ab}+c^\dagger_b|GS\rangle_{aa} \right)~.
\end{eqnarray*}
The highest weight state of the antisymmetric multiplet $|GS+1\rangle^A$, which 
transforms as $[2S-1,1]$ is a state orthogonal to $|GS+1\rangle^{S}_{aab}$, that is 
\begin{eqnarray*}
|GS+1\rangle^{A}_{aab} = \frac{1}{\Lambda\sqrt{2S}} \left(c^\dagger_a|GS%
\rangle_{ab}-\sqrt{2S-1}~c^\dagger_{b}|GS\rangle_{aa} \right)~,
\end{eqnarray*}
with a different normalization factor, $\Lambda^2 = \frac{q-1}{N-1}$.
Other states can be obtained from these three. The results are summarized in
the Tables  \ref{t1}-\ref{t3} of Clebsch-Gordan coefficients.

Next, we have to add the $(n_d-1)$ electrons on site 1. The result of the calculations are summarized in the Tables \ref{bigt}. We have 
\begin{widetext}
\begin{eqnarray*}
&&   \!\!\!\!\!\!\!\!\!\!\!\!\!\!\!\!\!\!\!\!\!\!  
|GS+1,n_d-1\rangle^{S}_{aab\cdots u} = \frac{1}{\sqrt{2S+n_d-1}} 
\left(%
\sqrt{2S}~(\prod_{i=2}^{n_d} d^\dagger_{x_i}) |GS+1\rangle^{S}_{aaa}
+\sum_{j=2}^{n_d} (-1)^{j-1}(\prod_{i=1,i\neq j}^{n_d} d^\dagger_{x_i} )
|GS+1\rangle^{S}_{aax_j} \right)
\\
&& \!\!\!\!\!\!\!\!\!\!\!\!\!\!\!\!\!\!\!\!\!\!
= \frac{1}{\Omega \sqrt{2S(2S+{n_d}-1)}} \left( 2S~(\prod_{i=2}^{n_d}
d^\dagger_{x_i}) c^\dagger_a|GS\rangle_{aa} \right.  
+ \left. \sum_{j=2}^{n_d} (-1)^{j-1}(\prod_{i=1,i\neq j}^{n_d} d^\dagger_{x_i} ) (%
\sqrt{2S-1}~c^\dagger_a|GS\rangle_{ax_j}+c^\dagger_{x_j}|GS\rangle_{aa})
\right)~,
\end{eqnarray*}
for the symmetric state in the symmetric configuration, and 
\begin{eqnarray*}
\lefteqn{|\overline{GS+1,n_d-1}\rangle^{S}_{aab\cdots u} =\frac{1}{%
\sqrt{{n_d}-1}}\left(\sum_{j=2}^{{n_d}} (-1)^j (\prod_{i=1,i\neq j}^{n_d}
d^\dagger_{x_i}) |GS+1\rangle^{A}_{aax_j}\right) } \\
&=& \frac{1}{\Lambda\sqrt{2S ({n_d}-1)}}\left(\sum_{j=2}^{n_d}(-1)^j
(\prod_{i=1,i\neq j}^{n_d} d^\dagger_{x_i}) (c^\dagger_a|GS\rangle_{ax_j}-\sqrt{%
2S-1}~c^\dagger_{x_j}|GS\rangle_{aa}) \right)~,
\end{eqnarray*}
for the antisymmetric state in the symmetric configuration
($x_1=a$). From here, it is
easy to show that the effect of $(c^\dagger_\sigma d_\sigma)$ on the ground
state, is given by Eq. (\ref{symm}).

\subsection{Antisymmetric process, $|GS,n_d\rangle^{A}$}

The action of $(c^\dagger_\sigma d_\sigma)$ on the ground state $%
|GS,n_d\rangle^{A}$, with $n_d$ electrons on site 1 coupled antisymmetrically, produces 
\begin{eqnarray}  
\lefteqn{\left(\sum_\sigma c^\dagger_\sigma d_\sigma\right)
|GS,n_d\rangle^{A}_{ab\cdots v} = \frac{(-1)^{{n_d}-1}}{\sqrt{2S+{n_d}-1}}
\left( \sqrt{2S-1} \sum_{l=2}^{{n_d}+1}(-1)^{l} (\prod_{i=2,i\neq l}^{{n_d}+1}
d^{\dagger}_{i} )c^\dagger_{x_l}|GS\rangle_{aa} \right. } &&  \nonumber \\
&+& \left. \sum_{j=2}^{{n_d}+1} (-1)^{j-1} \left\{ \sum_{l=1}^{j-1} (-1)^{l-1}
(\prod_{i=1,i\neq j,l}^{{n_d}+1} d^\dagger_i) +\sum_{l=j+1}^{{n_d}+1} (-1)^{l}
(\prod_{i=1,i\neq j,l}^{{n_d}+1} d^\dagger_i) \right\}
c^\dagger_{x_l}|GS\rangle_{ax_j} \right)~,  \label{kanti}
\end{eqnarray}
which is proportional to a given strong coupling excited state. Since we can
write (\ref{kanti}) as 
\begin{eqnarray*}
\left(\sum_\sigma c^\dagger_\sigma d_\sigma\right)
|GS,n_d\rangle^{A}_{ab\cdots v} &=& \frac{(-1)^{{n_d}-1}}{\sqrt{2S+{n_d}-1}} 
\left( \sum_{l=2}^{{n_d}+1}(-1)^{l}(\prod_{i=2,i\neq l}^{{n_d}+1}
d^\dagger_{x_i})(\sqrt{2S-1}~c^\dagger_{x_l}|GS\rangle_{aa}
-c^\dagger_a|GS\rangle_{ax_l}) \right.  \\
&& \!\!\!\!\!\!\!\!\!\!\!\!\!\!\!\!\!\!\!\!\!\!\!\!
+ \left. \sum_{j=2}^{{n_d}+1}(-1)^{j-1} \left(
\sum_{l=2}^{j-1}(-1)^{l-1}(\prod_{i=1,i\neq j,l}^{{n_d}+1} d^\dagger_{x_i}) -
\sum_{l=j+1}^{{n_d}+1}(-1)^{l-1}(\prod_{i=1,i\neq j,l}^{{n_d}+1}
d^\dagger_{x_i})\right) c^\dagger_{x_l}|GS\rangle_{ax_j}\right)  \\
\end{eqnarray*}
after some algebra, we get 
\begin{eqnarray*}
\left(\sum_\sigma c^\dagger_\sigma d_\sigma\right)
|GS,n_d\rangle^{A}_{ab\cdots v} &=& \frac{(-1)^{{n_d}-1}}{\sqrt{2S+{n_d}-1}}
\sum_{l=2}^{{n_d}+1}(-1)^{l} 
\left\{ (\prod_{i=2,i\neq l}^{{n_d}+1} d^\dagger_{x_i})(\sqrt{2S-1}%
~c^\dagger_{x_l}|GS\rangle_{aa} -c^\dagger_a|GS\rangle_{ax_l}) \right. \\
&-& \left. \sum_{j=2}^{l-1}(-1)^j(\prod_{i=1,i\neq j,l}^{{n_d}+1}
d^\dagger_{x_i}) (c^\dagger_{x_l}|GS\rangle_{ax_j}-
c^\dagger_{x_j}|GS\rangle_{ax_l}) \right\}~.
\end{eqnarray*}
This expression can be written using the antisymmetric states 
$|GS+1\rangle^{A}$, with the help of the following relations 
\begin{eqnarray*}
\Lambda(\sqrt{2S}~|GS+1\rangle^{A}_{ax_l x_j}- \sqrt{2S-2}%
~|GS+1\rangle^{A}_{ax_j x_l})= \sqrt{2S-1}(c^\dagger_{x_l}|GS%
\rangle_{ax_j}- c^\dagger_{x_j}|GS\rangle_{ax_l})~,
\end{eqnarray*}
\begin{eqnarray*}
\Lambda\sqrt{2S}~|GS+1\rangle^{A}_{aax_l}= c^\dagger_a|GS\rangle_{ax_l}-%
\sqrt{2S-1}~c^\dagger_{x_l} |GS\rangle_{aa}~,
\end{eqnarray*}
to obtain 
\begin{eqnarray*}
\lefteqn{\left(\sum_\sigma c^\dagger_\sigma d_\sigma\right)
|GS,n_d\rangle^{A}_{ab\cdots v} = \frac{(-1)^{{n_d}-1}\Lambda}{\sqrt{%
(2S+{n_d}-1)(2S-1)}}} && \\ && \times 
\sum_{l=1}^{{n_d}+1}(-1)^{l+1} 
\left\{ (\prod_{i=2,i\neq l}^{{n_d}+1} d^\dagger_{x_i})\sqrt{2S(2S-1)}%
~|GS+1\rangle^{A}_{aax_l} \right.  \ \\
&& - \left. \sum_{j=2}^{l-1}(-1)^j(\prod_{i=1,i\neq j,l}^{{n_d}+1}
d^\dagger_{x_i}) (\sqrt{2S}~|GS+1\rangle^{A}_{ax_jx_l}- \sqrt{2S-2}%
~|GS+1\rangle^{A}_{ax_lx_j}) \right\} ~. 
\end{eqnarray*}
Finally, since 
\begin{eqnarray}
\left(\sum_\sigma c^\dagger_\sigma d_\sigma\right)
|GS,n_d\rangle^{A}_{ab\cdots v} \propto
|GS+1,n_d-1\rangle^{A}_{ab\cdots v}
\end{eqnarray}
we just have to normalize the previous state in order to obtain the excited
state $|GS+1,n_d-1\rangle^{A}_{ab\cdots v}$. Up to a sign, we have 
\begin{eqnarray*}
\lefteqn{|GS+1,n_d-1\rangle^{A}_{ab\cdots v} = \frac{1}{\sqrt{%
{n_d}(2S+{n_d}-1)(2S-1)}}} && \\
&& \times \sum_{l=1}^{{n_d}+1}(-1)^{l+1} 
 \left\{ (\prod_{i=2,i\neq l}^{{n_d}+1} d^\dagger_{x_i})\sqrt{2S(2S-1)}%
~|GS+1\rangle^{A}_{aax_l} \right.  \\
&&- \left. \sum_{j=2}^{l-1}(-1)^j(\prod_{i=1,i\neq j,l}^{{n_d}+1}
d^\dagger_{x_i}) (\sqrt{2S}~|GS+1\rangle^{A}_{ax_jx_l}- \sqrt{2S-2}%
~|GS+1\rangle^{A}_{ax_lx_j}) \right\}~.
\end{eqnarray*}
and 
\begin{eqnarray*}
\left(\sum_\sigma c^\dagger_\sigma d_\sigma\right)
|GS,n_d\rangle^{A}_{ab\cdots v} = (\Lambda\sqrt{{n_d}})
|GS+1,n_d-1\rangle^{A}_{ab\cdots v} ~.
\end{eqnarray*}

\end{widetext}

\subsection{Tables of Clebsch-Gordan coefficients}

The calculation of the excited states involves the use of some
Clebsch-Gordon coefficients. We have evaluated these quantities explicitely
for arbitrary $2S$, $n_d$ and $N$, following the steps outlined in
Appendix \ref{Lshape}. We summarize our results in Tables \ref{t1}-\ref{bigt}.
In the tables, all the coefficients are assumed
to be under the sign of the square root. For instance, $-(2S-1)$ corresponds
to $-\sqrt{2S-1}$. The states should be divided by the normalization factor 
$\sqrt{N}$. Tables \ref{t1}-\ref{t3} are necessary when one electron 
is added to
the effective impurity, ($[1]\otimes[2S-1]$). Table \ref{bigt} corresponds to
the addition of $n_d$ electrons.
\begin{table}[h]
\caption {Clebsch-Gordan coefficients for the process $[1]\otimes [2S-1]\rightarrow
[2S]\oplus [2S-1,1]$. The label $|\{a\}aab\rangle$ indicates a state of $[2S]$,
whereas $|\{a\}aa,b\rangle$ denotes the highest weight state in $[2S-1,1]$
\label{t1}}
\begin{ruledtabular}
\begin{tabular}{c|c|cc}
&  &  &  \\ 
$aab$ & N & $|a\rangle|\{a\}ab\rangle$ & $|b\rangle|\{a\}aa\rangle$ \\ 
&  &  &  \\ \hline
&  &  &  \\ 
$|\{a\}aab\rangle$ & $2S$ & $2S-1$ & $1$ \\ 
&  &  &  \\ 
$|\{a\}aa,b\rangle$ & $2S$ & $1$ & $-(2S-1)$ \\ 
&  &  &  \\ 
\end{tabular}%
\end{ruledtabular}
\end{table}
\begin{table}[h]
\caption {Same as Table \protect\ref{t1}, but for states with quantum numbers 
$(a)^{2S-2}bb$
\label{t2}}
\begin{ruledtabular}
\begin{tabular}{c|c|cc}
&  &  &  \\ 
$abb$ & N & $|a\rangle|\{a\}bb\rangle$ & $|b\rangle|\{a\}ab\rangle$ \\ 
&  &  &  \\ \hline
&  &  &  \\ 
$|\{a\}abb\rangle$ & $2S$ & $2S-2$ & $2$ \\ 
&  &  &  \\ 
$|\{a\}ab,b\rangle$ & $2S$ & $2$ & $-(2S-2)$ \\ 
&  &  &  \\ 
\end{tabular}%
\end{ruledtabular}
\end{table}
\begin{table}[h]
\caption {Same as Table \protect\ref{t1}, but for states with quantum numbers $(a)^{2S-2}bc$.
Notice the degeneracy in $[2S-1,1]$. We denote corresponding orthogonal
states by $|\{a\}ab,c\rangle$, and $|\{a\}ac,b\rangle$.
\label{t3}}
\begin{ruledtabular}
\begin{tabular}{c|c|ccc}
&  &  &  &  \\ 
$abc$ & N & $|a\rangle|\{a\}bc]$ & $|b\rangle|\{a\}ac\rangle$ & $|c\rangle|\{a\}ab\rangle$ \\ 
&  &  &  &  \\ \hline
&  &  &  &  \\ 
$|\{a\}abc\rangle$ & $2S$ & $2S-2$ & $1$ & $1$ \\ 
&  &  &  &  \\ 
$|\{a\}ab,c\rangle$ & $2S(2S-1)$ & $2S-2$ & $1$ & $-(2S-1)^2$ \\ 
&  &  &  &  \\ 
$|\{a\}ac,b\rangle$ & $2S-1$ & $1$ & $-(2S-2)$ & $0$ \\ 
&  &  &  &  \\ 
\end{tabular}%
\end{ruledtabular}
\end{table} 
\begin{table*}[h]
\caption {Some of the CG coefficients for the product 
of $k=n_d$ electrons, 
and an effective impurity $[2S-1]$, ($[1^{k}]\otimes [2S-1]
\rightarrow [2S,1^{k-1}]\oplus [2S-1,1^k] \oplus \cdots$). Here we only 
keep the coefficients for the representations $[2S,1^{k-1}]$, 
and $[2S-1,1^k]$ (last row). 
States from $[1^{k}]$ are denoted by
a column of labels, $|\vdots\rangle$; those from $[2S-1]$ are
denoted $|\alpha\rangle$, and are states {\em close} to
the highest weight state, denoted $|a\rangle$. The states in 
$[2S,1^{k-1}]$ are labeled $|ab,c,d,\dots\rangle$ and those in 
$[2S-1,1^{k}]$, $|a,b,c,\cdots\rangle$ \label{bigt} }
\begin{ruledtabular}
\begin{tabular}{c|c|cccccc}
&&&&&&& \\
$\begin{array}{c} [1^{k}]\otimes [2S-1] \\ \\
\{a\}abc\cdots uv  \end{array}$
&$N$ & 
$[1^k]\{ |\vdots\rangle  
\overbrace{|a\rangle}^{[2S-1]}$ 
&
$|\vdots\rangle|b\rangle$ 
& 
$|\vdots\rangle|c\rangle$ 
& 
$\cdots$
&
$|\vdots\rangle|u\rangle$ 
& 
$|\vdots\rangle|v\rangle$  
\\
 &&&&&&& \\
\hline
&&&&&&& \\
$|ab,c,\cdots,u,v\rangle$ & $2S(2S-1)$ &
$2S-1$ & 
$(2S-1)^2$ & ${0}$ & ${\cdots}$ & ${0}$
 & ${0}$ \\ 
&&&&&&& \\
$|ac,b,\cdots,u,v\rangle$ & ${2S(2S+1)}$ &
${-(2S-1)}$ & ${1}$ & ${(2S)^2}$ 
& ${\cdots}$ & ${0}$ & ${0}$ \\
&&&&&&& \\
${\cdots}$ &  & ${\cdots}$ & ${\cdots}$ & 
${\cdots}$ & ${\cdots}$ & ${\cdots}$ 
& ${\cdots}$ \\
&&&&&&& \\ 
${|au,b,\cdots,t,v\rangle}$ & 
${(2S+k-2)(2S+k-3)}$ &
${(-1)^{k}(2S-1)}$ & ${-(-1)^{k}}$ & ${(-1)^{k}}$ 
& ${\cdots}$ & 
${(2S+k-3)^2}$ & ${0}$ \\
&&&&&&& \\
${|av,b,\cdots,t,u\rangle}$ & 
${(2S+k-1)(2S+k-2)}$ &
${(-1)^{k+1}(2S-1)}$ & ${-(-1)^{k+1}}$ & ${(-1)^{k+1}}$ 
& ${\cdots}$ & ${1}$ 
& ${(2S+k-2)^2}$ \\
&&&&&&& \\
\hline &&&&&&& \\  
${|a,b,\cdots,u,v\rangle}$ &
${(2S+k-1)}$ & 
${(2S-1)}$ & ${-1}$ & ${1}$ & ${\cdots}$ 
& ${(-1)^{k-1}}$ &
${(-1)^k}$ \\
&&&&&&& \\
\end{tabular}
\end{ruledtabular}
\end{table*}

\bibliography{biblio3}

\end{document}